\documentclass[prd,showpacs,twocolumn,superscriptaddress,nofootinbib,floatfix,10pt]{revtex4-2}
\usepackage{setspace}
\usepackage[utf8]{inputenc}
\usepackage{float}
\usepackage{dcolumn}
\usepackage{booktabs}
\usepackage{array}
\usepackage{amsmath,amssymb,amsfonts,bm}
\usepackage{graphicx}
\usepackage{multirow}
\usepackage{subfigure}
\usepackage{makecell}
\usepackage[usenames,dvipsnames]{color}
\usepackage{overpic}
\usepackage{verbatim}
\usepackage{ulem}

\allowdisplaybreaks
\usepackage[
colorlinks=true,
linkcolor=blue,
breaklinks=true,
urlcolor=blue,
citecolor=blue]{hyperref}

\definecolor{green}{rgb}{0,0.6,0}

\newcommand{\be}{\begin{equation}} 
\newcommand{\ee}{\end{equation}}
\newcommand{\bea}{\begin{eqnarray}} 
\newcommand{\eea}{\end{eqnarray}}
\newcommand{\beas}{\begin{eqnarray*}} 
\newcommand{\eeas}{\end{eqnarray*}}

\renewcommand{\vec}{\bm}

\newcommand{\moe}{\affiliation{Key Laboratory of Atomic and Subatomic Structure and Quantum Control (MOE), Guangdong Basic Research Center of Excellence for Structure and Fundamental Interactions of Matter, Institute of Quantum Matter, South China Normal University, Guangzhou 510006, China
}}

\newcommand{\ihep}{\affiliation{Institute of High Energy Physics, Chinese Academy of Sciences, Beijing 100049, China}}

\newcommand{\iqm}{\affiliation{Guangdong-Hong Kong Joint Laboratory of Quantum Matter, Guangdong Provincial Key Laboratory of Nuclear Science, Southern Nuclear Science Computing Center, South China Normal University, Guangzhou 510006, China}}

\newcommand{\scnt}{\affiliation{Southern Center for Nuclear-Science Theory (SCNT), Institute of Modern Physics, Chinese Academy of Sciences, Huizhou 516000, Guangdong Province, China}}

\newcommand{\cas}{\affiliation{CAS Key Laboratory of Theoretical Physics, Institute of Theoretical Physics, Chinese Academy of Sciences, Beijing 100190, China}}

\newcommand{\casu}{\affiliation{University of Chinese Academy of Sciences, Beijing 100049, China}}

\begin{document}

\title{Revealing the mystery of the double charm tetraquark in $pp$ collision}

\author{Xue-Li Hua}
\email{xueli.hua@m.scnu.edu.cn}
\moe\iqm

\author{Yi-Yao Li}
\email{liyiyao@m.scnu.edu.cn, co-first author}
\moe\iqm

\author{Qian Wang}
\email{qianwang@m.scnu.edu.cn}
\moe\iqm\scnt

\author{Shuai Yang}
\email{syang@scnu.edu.cn}
\moe\iqm

\author{Qiang Zhao}
\email{zhaoq@ihep.ac.cn}
\ihep\casu

\author{Bing-Song Zou}
\email{zoubs@itp.ac.cn}
\cas\casu\scnt

\begin{abstract}
 A novel approach is proposed to probe the nature of the double charm tetraquark through the prompt production asymmetry between $T_{\bar{c}\bar{c}}^-$ and $T_{cc}^+$ in $pp$ collisions. When comparing two theoretical pictures, i.e. the compact tetraquark and hadronic molecular pictures, we find that the former one exhibits a significantly larger production asymmetry, enabling the unambiguous determination of the tetraquark's internal structure. Additionally, distinctive differences in the transverse momentum and rapidity distributions of $T_{\bar{c}\bar{c}}^-$ and $T_{cc}^+$ cross sections emerge, particularly at $p_\mathrm{T}\approx 2~\mathrm{GeV}$ and $y\approx \pm 6$ at a center-of-mass energy of 14$~\mathrm{TeV}$.  
 The insignificant asymmetry in hadronic molecular picture is because that hadronic molecules are produced in hadronic phase,
 where the phase space of their constituents needs to be taken into account rigorously. Our work can be extended to the exploration of other double heavy tetraquark candidates, offering a versatile approach to advance our understanding of exotic hadrons.
\end{abstract}

\maketitle

\section{Introduction}

 Quantum Chromodynamics (QCD) is accepted as the fundamental theory of strong interaction, which exhibits color confinement, resulting in the experimental observation of color-singlet hadrons. They can be classified into mesons, which are composed of quark-antiquark pairs, and baryons, which consist of three quarks in the conventional quark model. Exotic hadrons are color-singlets with constituent structures beyond the scenarios of the conventional quark model.  
During the past two decades, tens of exotic candidates have been observed due to the significant progress in experiment~\cite{Olsen:2014qna}. In particular, a large number of exotic candidates are located in charmonium energy region including the hidden charm pentaquarks $P_c$~\cite{LHCb:2019kea,LHCb:2015yax}, the hidden charm tetrquarks $Z_c^{(\prime)}$~\cite{BESIII:2013ris,Belle:2013yex,Xiao:2013iha,BESIII:2013ouc}, $Z_{cs}$\cite{BESIII:2020qkh,LHCb:2021uow},
the double charm tetraquarks $T_{cc}^+$~\cite{LHCb:2021vvq,LHCb:2021auc}, and so on. 
Among these candidates, the double charm tetraquark $T_{cc}^+$ observed in the decay channel $D^0D^0\pi^+$ by the LHCb Collaboration is of great interest. Since $T_{cc}^+$ must contain two charm quarks and two light antiquarks, it does not mix with normal mesons and is regarded as a perfect candidate for an exotic hadron.

To date, various of theoretical interpretations have been proposed for these exotic candidates, such as compact tetraquarks~\cite{Esposito:2014rxa,Lebed:2016hpi}, hadronic molecules~\cite{Guo:2017jvc}, hybrids~\cite{Meyer:2015eta}, normal heavy quarkonium~\cite{QuarkoniumWorkingGroup:2004kpm}, and so on.  For detailed reviews, one can refer to Refs.~\cite{Esposito:2014rxa,Lebed:2016hpi,Guo:2017jvc,Chen:2016qju,Yamaguchi:2019vea,Brambilla:2019esw,Guo:2019twa,Chen:2022asf,Kalashnikova:2018vkv,Meyer:2015eta}. In order to further deepen the understanding of exotic hadrons, physicists are putting tremendous effort into exploring novel research methods and topics. This includes exploring the properties of exotic hadrons using machine learning techniques ~\cite{Zhang:2023czx,Liu:2022uex,Niu:2018csp,Niu:2018trk,Ma:2020mbd,Kaspschak:2020ezh,Kaspschak:2021hbc,Bedaque:2021bja,Dong:2022wkd}, and studying their behavior in nuclear matter~\cite{Guo:2023dwf,Montesinos:2023qbx,Albaladejo:2021cxj,Cleven:2019cre,Carames:2016qhr,Abreu:2016qci}, as well as their production in heavy ion collisions~\cite{Zhang:2020dwn,Hu:2021gdg,ExHIC:2010gcb,Chen:2023xhd,Yun:2022evm,Chen:2021akx,Wu:2020zbx,MartinezTorres:2014son,Cho:2013rpa,ExHIC:2013pbl,ExHIC:2011say}.
Despite the abundance of theoretical interpretations regarding the properties of exotic candidates, controversial interpretations are often seen in the literature.
 Taking the double charm tetraquark $T_{cc}^+$ as an example, 
both the compact tetraquarks ~\cite{Meng:2023for,Noh:2023zoq,Song:2023izj,Wu:2022gie,Simonov:2022bdy,Kim:2022mpa,Weng:2021hje,Gao:2020bvl} and hadronic molecular pictures ~\cite{Shi:2022slq,Albaladejo:2021vln,Vijande:2003ki,Ebert:2007rn,Karliner:2017qjm,Eichten:2017ffp,Ader:1981db,Luo:2017eub,Liu:2023hrz,Wang:2022jop,Ortega:2022efc,Chen:2022vpo,Lin:2022wmj,Mikhasenko:2022rrl,Ke:2022vsi,Padmanath:2022cvl,Agaev:2022ast,Ke:2021rxd,Bilmis:2021rdp,Deng:2021gnb,Du:2021zzh,Chen:2021cfl,Xin:2021wcr} can describe its mass position, but with different underlying physics.

In this work, we explore the prompt production of the double charm tetraquark $T_{cc}^+$ and its antimatter counterpart $T_{\bar{c}\bar{c}}^-$ in $pp$ collision at a center-of-mass energy $\sqrt{s} = 14~\mathrm{TeV}$. 
Our investigation reveals that the production asymmetry between $T_{\bar{c}\bar{c}}^-$ and $T_{cc}^+$ serves as a crucial and directly measurable physical quantity, offering insights into the intrinsic characteristics of the double charm tetraquark. Furthermore, notable distinctions manifest in the transverse momentum ($p_\mathrm{T}$) and rapidity ($y$) distributions of $T_{\bar{c}\bar{c}}^-$ and $T_{cc}^+$ cross sections. It is worth noting that this analytical approach can be extended to the investigations of other double heavy tetraquark candidates. 

\vspace{0.2cm}
\section{Framework}

The $T_{cc}^+$ is observed in the $D^0D^0\pi^+$~\cite{LHCb:2021vvq,LHCb:2021auc} 
invariant mass distribution and its mass is very close to the $DD^{\ast}$ threshold. The absence of a peak in the $D^+D^+$ invariant mass distribution~\cite{LHCb:2021vvq} suggests that it could be an isospin singlet $DD^{\ast}$ hadronic molecule candidate~\cite{Qiu:2023uno,Du:2021zzh,Chen:2023fgl,Wang:2023ovj,He:2023ucd,Peng:2023lfw,Dai:2023cyo,Meng:2021jnw,Lin:2022wmj,Pan:2022xxz,Kamiya:2022thy,Ke:2021rxd,Deng:2021gnb,Dai:2021vgf,Vidana:2023olz,Jia:2022qwr}. It can also be accepted as a compact tetraquark with $cc$ diquark and $\bar{u}\bar{d}$ antidiquark with respect to the experimental mass position~\cite{Meng:2023for,Noh:2023zoq,Song:2023izj,Wu:2022gie,Simonov:2022bdy,Kim:2022mpa,Weng:2021hje,Gao:2020bvl,Galkin:2023wox,Meng:2023jqk,Ma:2023int,Yang:2009zzp,Deng:2022cld,Vijande:2009ac,Anwar:2023svj}. Consequently, the production of double charm tetraquarks involves a two-step process: the generation of their constituents  and the subsequent formation of the double charm tetraquarks. In our framework, the 
PYTHIA 8.3~\cite{Bierlich:2022pfr} is employed to simulate $pp$ collisions at $\sqrt{s} = 14~\mathrm{TeV}$, corresponding the top operational energy of the Large Hadron Collider (LHC).

In the molecular picture,
we first simulate the prompt production of the $DD^{\ast}$ and $\bar{D}\bar{D}^{\ast}$ pairs for the formation of the $T_{cc}^+$ and $T_{\bar{c} \bar{c}}^-$, respectively. 
 The formation of the $T_{cc}^+$ is based on the factorization firstly proposed in Refs.~\cite{Artoisenet:2009wk,Artoisenet:2010uu} and
further improved in Ref.~\cite{Guo:2013ufa}, which considers phase spaces of $D$ and $D^{\ast}$ in the $DD^{\ast}$ pair production. This has been successfully applied to the production of the $X(3872)$ ~\cite{Guo:2014sca,Albaladejo:2017blx,Yang:2021jof,Shi:2022ipx}, $P_c$s~\cite{Ling:2021sld}, $Z_b$s~\cite{Cao:2019gqo,Guo:2013ufa}, $D_{s0}(2317)$~\cite{Guo:2014ppa} and their heavy quark spin partners. More specifically, taking the $T_{cc}^+$ as an example, the production cross section can be written as
\begin{equation}
\sigma(T_{cc})=\frac{1}{4m_Dm_{D^{\ast}}}g^2|\mathcal{G}|^2(\frac{d\sigma(DD^{\ast})}{dk})\frac{4\pi^2\mu}{k^2}\label{fomula:6}
\end{equation} 
where the $m_{D}$, $m_{D^*}$ and $\mu$ are the $D$ meson mass, the $D^*$ meson mass and their reduced mass, respectively. $k=\sqrt{2\mu (E-m_D-m_{D^\ast})}$ and $E$ are the relative three momentum in the $DD^*$ rest frame and their total energy, respectively. The $\frac{d\sigma(DD^{\ast})}{dk}$ is the $DD^\ast$ pair differential cross section, which is proportional to $k^2$ \cite{Guo:2014sca},  reflecting the phase space factor. $\mathcal{G}$ is the $DD^\ast$ two-body propagator and linearly divergent \cite{Guo:2014sca}\footnote{The production of hadronic molecule in hadron collision was studied by convoluting the cross sections of the constituents and the wave function in  Refs.~\cite{Bignamini:2009sk,Esposito:2017qef}. However, the momentum integrated region is still under debate~\cite{Albaladejo:2017blx}, leaving the cross section dependent on the integrated upper limit. The wave function in Refs.~\cite{Albaladejo:2017blx,Bignamini:2009sk,Esposito:2017qef} plays the same role as the two-body propagator in our framework. At the same time, the momentum integrated upper limit~\cite{Albaladejo:2017blx,Bignamini:2009sk,Esposito:2017qef} is similar to the role of $\Lambda$ in our framework, 
i.e. controlling the long-ranged contribution. 
Although the cross sections do depend on the choice of the two-body propagator and the value of $\Lambda$,
the asymmetries of various choices are consistent with each other within uncertainties. That makes the asymmetry a crucial quantity for the internal structure of double charm tetraquark. This is the important message that we want to deliver in our manuscript. 
}
\begin{eqnarray}
\mathcal{G}(E) &\equiv& \int_{0}^{\Lambda}\frac{q^2dq}{2\pi^2} \frac{2\mu}{q^2-k^2-i\epsilon} \\
&=&\frac{\mu \Lambda}{\pi^2}-\frac{\mu k}{\pi^2}\mathrm{ArcTanh}(\frac{k}{\Lambda})+i\frac{\mu k}{\pi},\label{fomula:2}
\end{eqnarray}

The effective coupling constant $g$ is defined as \cite{Guo:2014sca}
\begin{equation}
g^2\equiv\lim_{E \to m_{T_{cc}}}(E^2-m_{T_{cc}}^2)\frac{C(\Lambda)}{1-C(\Lambda)\mathcal{G}(E,\Lambda)}.\label{fomula:7}
\end{equation}
 $C$ is the leading order low energy constant which describes the contact interaction of the $DD^{\ast}$ scattering. The bound state pole satisfies $1-C\mathcal{G}[E_{\mathrm{pole}}]=0$ in the leading order. One notice that our framework for molecular picture is different from the coalescence model in Refs.~\cite{Esposito:2014rxa,Guerrieri:2014gfa,Esposito:2016noz},
where the phase space factor is considered as a constant and reabsorbed into the normalization factor fixed by comparing with the experimental data.

  In the compact tetraquark picture, color and spin information is not taken into account, given that PYTHIA 8.3 only provides color flow information without specifying exact color details. We gathered all the generated $c$ quarks and $\bar{c}$ quarks within a single event, retaining only (anti)charmed diquarks with a relative three-momentum less than $1~\mathrm{GeV}$, as an order-of-magnitude estimate, for $cc$ and $\bar{c}\bar{c}$ pairs. The same method 
  is also used to obtain light diquark $ud$ and anti-diquark $\bar{u}\bar{d}$. 
  Besides those, the $ud_0$ and $ud_1$ diquarks from Monte Carlo 
  simulation are also considered. The subindices mean their total spin. To take into account the formation dynamics, we introduce a Gaussian function 
   \begin{equation}
 \phi(\vec{p},a)=(\frac{1}{\pi})^{\frac{3}{4}}(\frac{1}{a})^{\frac{3}{2}}e^{\frac{-p^2}{2a^2}},\label{eq:wf1}
 \end{equation}
 in momentum space between diquark and antidiquark,
 with $\vec{p}$ their relative three momentum and $p$ its absolute value.
The normalization of this wave function read as
\begin{equation}
    \int |\phi(\vec{p},a)|^2 \mathrm{d}^3\vec{p}=1.
\end{equation}
The corresponding wave function
\begin{equation}
    \phi_\mathrm{c}(\vec{r},a)=(\frac{1}{\pi})^{\frac{3}{4}}a^{\frac{3}{2}}e^{\frac{-a^2 r^2}{2}}
\end{equation}
in coordinate space can be obtained by a Fourier Transformation with $r=|\vec{r}|$.
The mean square radius
\begin{eqnarray}
    \langle r \rangle=\sqrt{\int |\phi_\mathrm{c}(\vec{r},a)|^2 r^2 \mathrm{d}^3\vec{r}}=\sqrt{\frac{3}{2}}\frac{1}{a} \label{eq:r}
\end{eqnarray}
reflects the size of the compact tetraquark. 
To analyse systematic uncertainty from wave function, 
we also introduce a Sturmian type function
\begin{equation}
 \psi(\vec{p},b)=2\sqrt{2}(\frac{1}{\pi})(\frac{1}{b})^{\frac{3}{2}}\frac{1}{(\frac{p^2}{b^2}+1)^2},\label{eq:wf2}
 \end{equation}
 as a comparison. 
 The corresponding wave function in coordinate space is
 \begin{equation}
     \psi_\mathrm{c}(\vec{r},b)=\frac{1}{\sqrt{\pi}} b^{\frac{3}{2}} e^{-br}.
 \end{equation}
 The corresponding mean square radius is $\langle r \rangle=\frac{\sqrt{3}}{b}$. 
The systematic uncertainty associated to the hadron size is determined by altering the mean square radius from $0.2~\mathrm{fm}$ to $1~\mathrm{fm}$.

\begin{table*}[htbp]
\centering
\caption{The cross sections of the $T_{cc}^{+}$ (without brackets), $T_{\bar{c} \bar{c}}^{-}$ (with brackets) and their asymmetry $\mathcal{A}$ (the fifth column) for the molecular picture in $pp$ collisions at $\sqrt{s} = 14~\mathrm{TeV}$. The second, third and fourth column represent three different choices of $\Lambda$ (Eq.~\eqref{fomula:2}), i.e. $0.5~\mathrm{GeV}$, $1~\mathrm{GeV}$ and $1.5~\mathrm{GeV}$. 
The second row represents the cross sections and the asymmetry without any kinematic constraint, while the third, fourth, and fifth rows are for those with the LHCb, CMS and ATLAS kinematic acceptance, respectively. The $p_\mathrm{T}$ cut of $10~\mathrm{GeV}$ is chosen based on the experimental measurements of the $X(3872)$~\cite{LHCb:2021ten,LHCb:2021auc,CMS:2013fpt,ATLAS:2016kwu}. One should keep in mind that the cross sections are an order-of-magnitude estimate. 
The uncertainties are only the statistic ones. One can see a correlation between the cross sections of particles and those of antiparticles, leaving the asymmetry $\mathcal{A}$ as a model-independent measurable quantity for the property of double charm tetraquark. The first and second uncertainties of the asymmetry in the last column are the systematic and statistic ones, respectively.}
\label{table:cs-HM}
\begin{tabular}{m{11em}  m{3cm}  m{3cm} m{3cm} m{3cm}} 
  \hline\hline
  \multirow{2}{11em}{Range($\mathrm{GeV}$)} & \multicolumn{3}{c}{\multirow{2}*{$\sigma_{T_{cc}^{+}}$($\sigma_{T_{\bar{c} \bar{c}}^{-}}$)}} & \multirow{2}{9em}{$\mathcal{A}(\%)$} \\
   \\ \hline
  &\makecell*[c]{$\Lambda=0.5~\mathrm{GeV}$}&\makecell*[c]{$\Lambda=1~\mathrm{GeV}$}&\makecell*[c]{$\Lambda=1.5~\mathrm{GeV}$}\\

  \hline\hline
  
  \multirow{3}{11em}{Full} &
  \makecell*[c]{$43.30\pm$0.70 $\mathrm{nb}$} & \makecell*[c]{$152.42\pm$0.89 $\mathrm{nb}$} & \makecell*[c]{$313.74\pm$1.03 $\mathrm{nb}$} & \multirow{3}{9em}{$1.24\pm$$0.30\pm$0.20 }\\ &
  \makecell*[c]{($44.13\pm$0.71 $\mathrm{nb}$)} & \makecell*[c]{($156.81\pm$0.91 $\mathrm{nb}$)} & \makecell*[c]{($321.14\pm$1.04 $\mathrm{nb}$)}\\ 
  
  \hline
  \specialrule{0em}{4pt}{0pt}
  \multicolumn{5}{c}{LHCb ($2<y<4.5$)}\\
  \specialrule{0em}{2pt}{0pt}

  \multirow{2}{11em}{$4<p_T<20$~\cite{LHCb:2021ten}} &
  \makecell*[c]{$1.46\pm$0.15 $\mathrm{nb}$} & \makecell*[c]{$5.27\pm$0.20 $\mathrm{nb}$} & \makecell*[c]{$11.46 \pm$0.23 $\mathrm{nb}$} & \multirow{3}{9em}{$2.53\pm$$2.01\pm$1.79 }\\ &
  \makecell*[c]{($1.45\pm$0.15 $\mathrm{nb}$)} & \makecell*[c]{($5.63\pm$0.20 $\mathrm{nb}$)} & \makecell*[c]{($11.87\pm$0.24  $\mathrm{nb}$)}\\ 
  
  \specialrule{0em}{5pt}{0pt}
  \multirow{2}{11em}{$p_T>0$~\cite{LHCb:2021auc}} &
  \makecell*[c]{$8.26 \pm$0.44 $\mathrm{nb}$} & \makecell*[c]{$29.93\pm$0.57 $\mathrm{nb}$} & \makecell*[c]{ $62.28\pm$0.66 $\mathrm{nb}$} & \multirow{3}{9em}{$1.64\pm$$1.03\pm$0.52}\\ &
  \makecell*[c]{($8.69\pm$0.46 $\mathrm{nb}$)} & \makecell*[c]{($30.82\pm$0.58 $\mathrm{nb}$)} & \makecell*[c]{($64.30\pm$ 0.67 $\mathrm{nb}$)} \\ 
  
  \hline
  \specialrule{0em}{4pt}{0pt}
   \multicolumn{5}{c}{CMS ($|y|<1.2$)} \\
   \specialrule{0em}{2pt}{0pt}
  \multirow{2}{11em}{$10<p_T<50(30)$~\cite{CMS:2013fpt}} &
  \makecell*[c]{$0.05\pm$0.02 $\mathrm{nb}$} & \makecell*[c]{$0.28\pm$0.04 $\mathrm{nb}$} & \makecell*[c]{$0.55\pm$0.04 $\mathrm{nb}$} & \multirow{3}{9em}{$-13.42\pm$$8.44\pm$2.18}\\ &
  \makecell*[c]{($0.03\pm$0.02 $\mathrm{nb}$)} & \makecell*[c]{($0.20\pm$0.03 $\mathrm{nb}$)} & \makecell*[c]{($0.44\pm$0.04 $\mathrm{nb}$)}\\ 
  \hline
  \specialrule{0em}{4pt}{0pt}
   \multicolumn{5}{c}{ATLAS ($|y|<0.75$)} \\
   \specialrule{0em}{2pt}{0pt}
   \multirow{2}{11em}{$10<p_T<70$~\cite{ATLAS:2016kwu}} &
  \makecell*[c]{$0.03\pm$0.02 $\mathrm{nb}$} & \makecell*[c]{$0.20\pm$0.03 $\mathrm{nb}$}  & \makecell*[c]{$0.38\pm$0.04 $\mathrm{nb}$ } & \multirow{3}{9em}{$-16.87\pm$$9.33\pm$10.10}\\ &
  \makecell*[c]{($0.03\pm$0.02 $\mathrm{nb}$)} & \makecell*[c]{($0.13\pm$0.03 $\mathrm{nb}$)} & \makecell*[c]{($0.28\pm$0.03 $\mathrm{nb}$)} \\ 
  
  \hline\hline
\end{tabular}
\end{table*}

\begin{figure*}[htbp]
\flushleft
\subfigure{
\begin{minipage}[b]{.3\linewidth}
\begin{overpic}[scale=0.29]{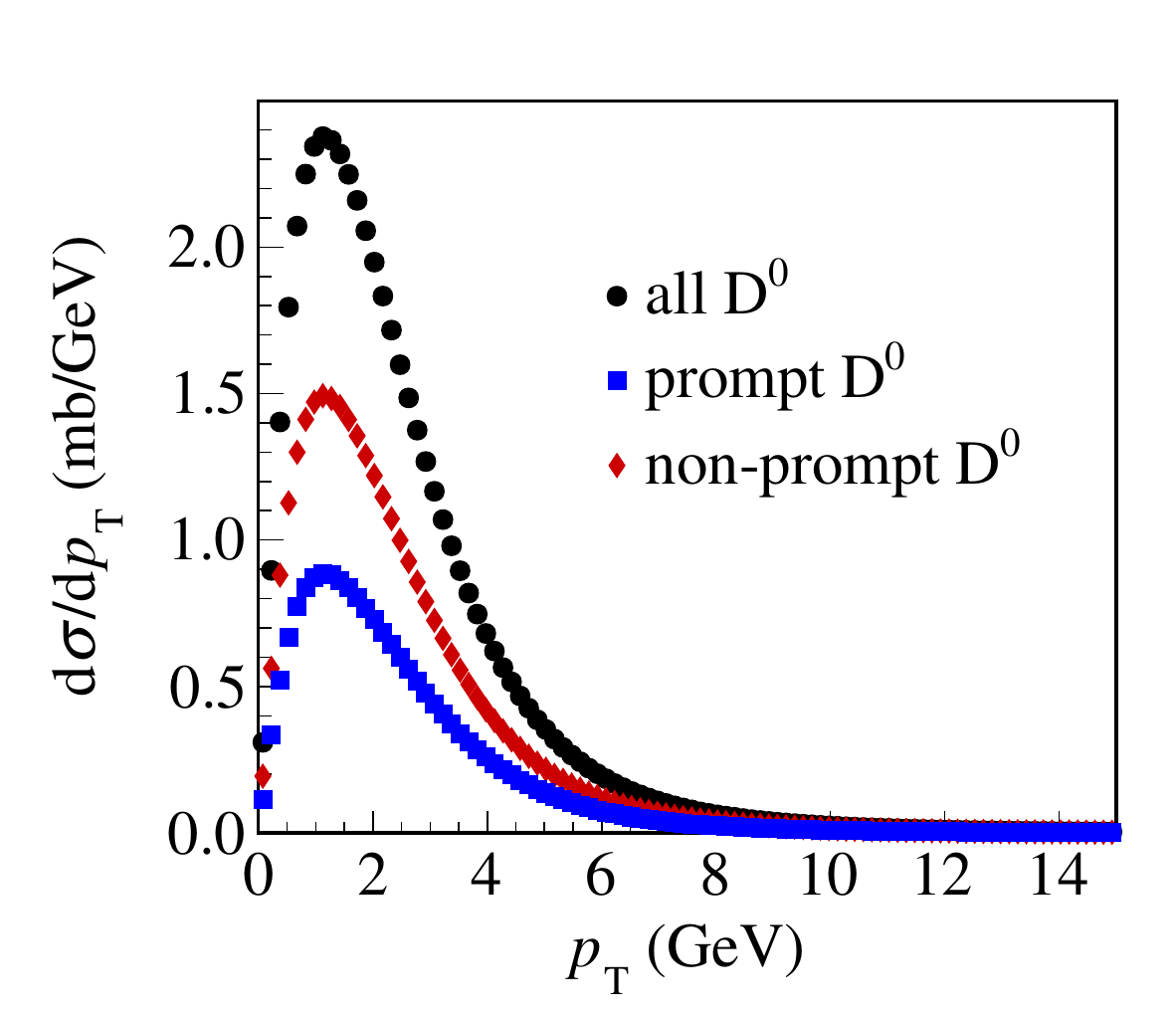}
 \put(3,75){$\textbf{(a)}$}
\end{overpic}
\end{minipage}
}
\subfigure{
\begin{minipage}[b]{.3\linewidth}
\begin{overpic}[scale=0.29]{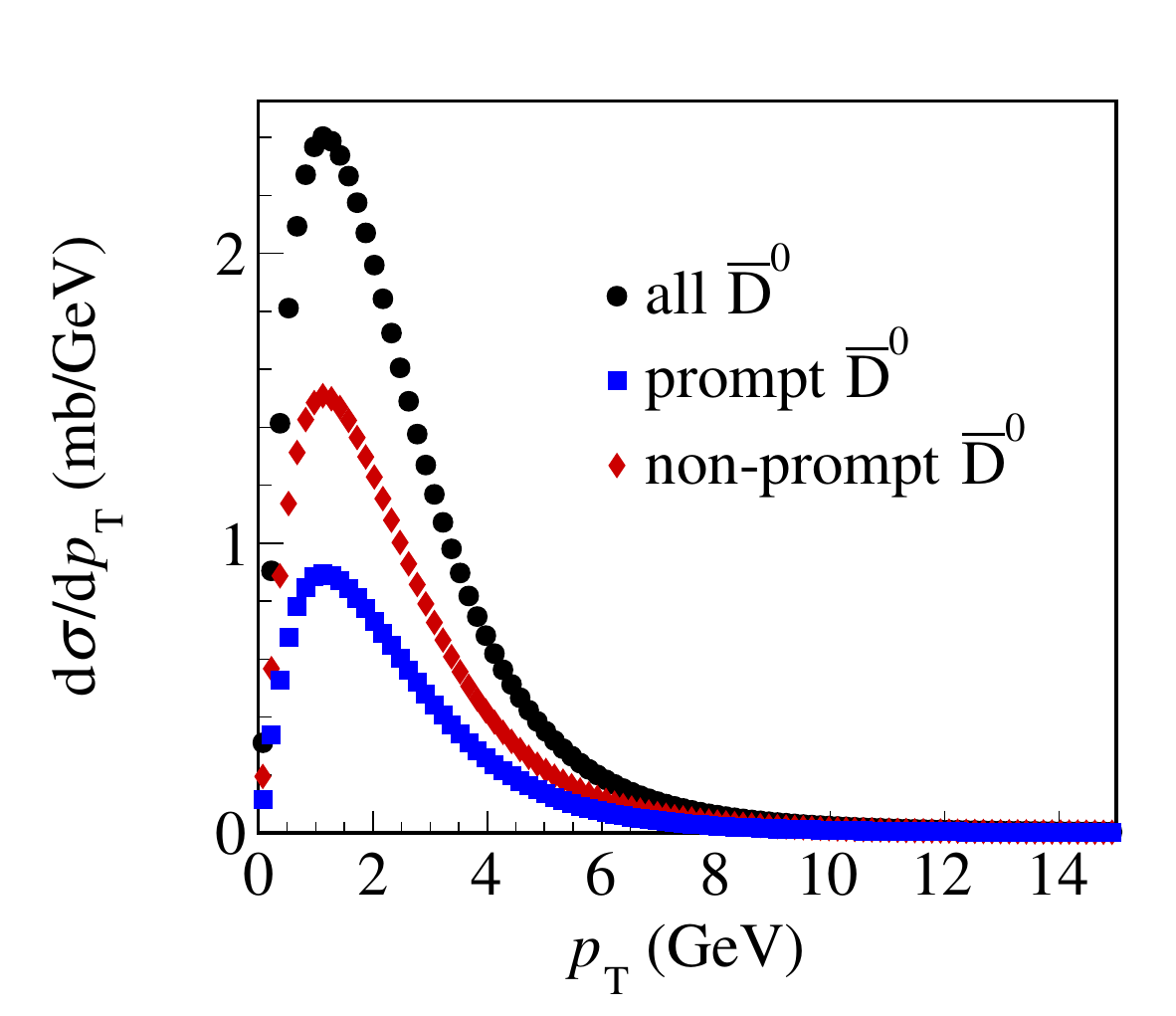}
 \put(3,75){$\textbf{(b)}$}
\end{overpic}
\end{minipage}
}
\subfigure{
\begin{minipage}[b]{.3\linewidth}
\begin{overpic}[scale=0.29]{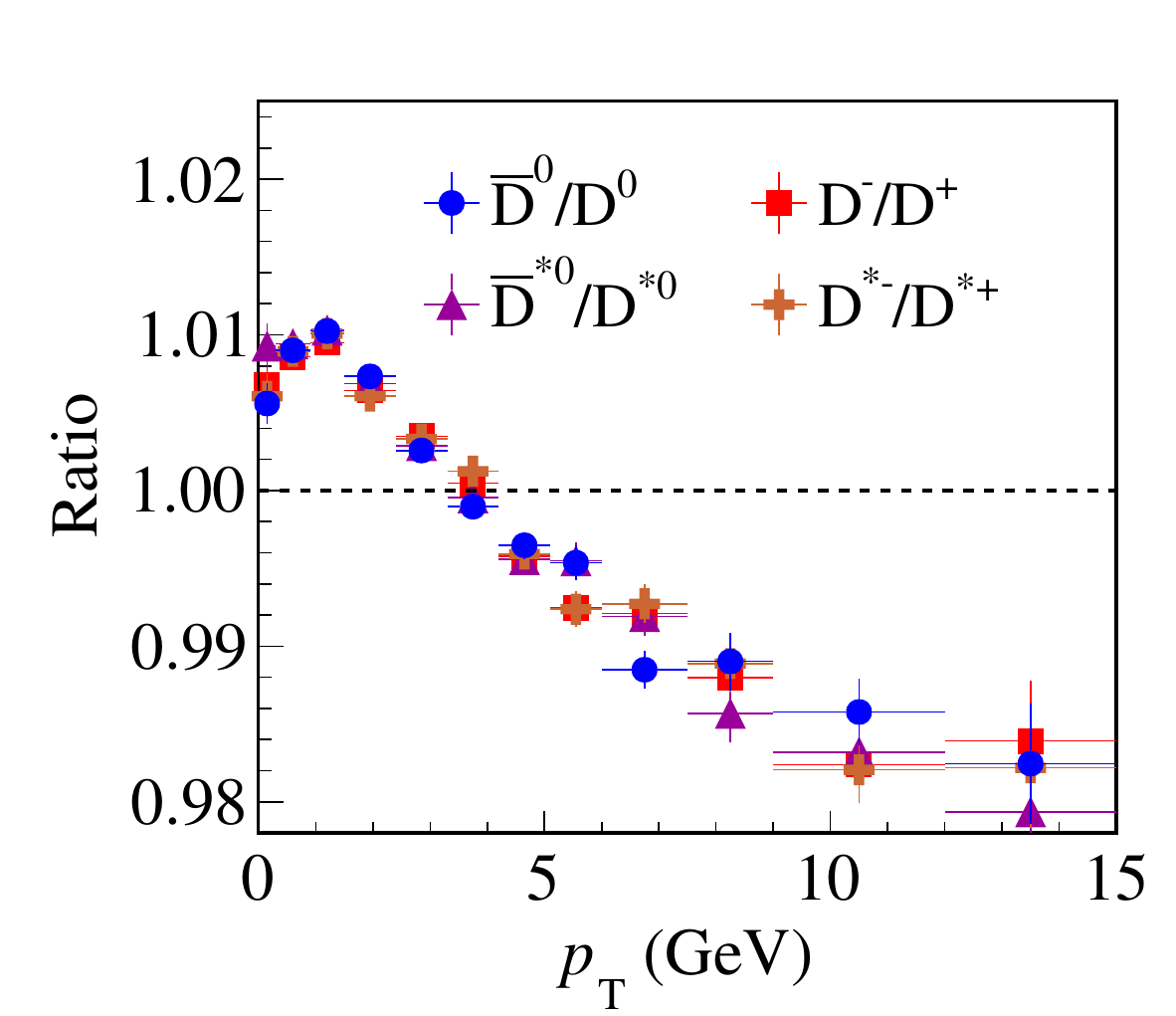}
 \put(3,75){$\textbf{(c)}$}
\end{overpic}
\end{minipage}
}
\caption{(a): The $p_\mathrm{T}$ distributions of the prompt $D^0$ (blue squares), non-prompt $D^0$ (red diamonds) and their combined distribution (black circles) in the $pp$ collision at $\sqrt{s} = 14~\mathrm{TeV}$. (b): The $p_\mathrm{T}$ distributions of the prompt $\bar{D}^0$ (blue squares), non-prompt $\bar{D}^0$ (red diamonds) and their combined distribution (black circles). (c): The prompt production ratios of the $\bar{D}^0/D^0$ (blue circles), $D^-/D^+$ (red squares), $\bar{D}^{*0}/D^{*0}$ (purple triangles), $D^{*-}/D^{*+}$ (orange crosses) as a function of $p_\mathrm{T}$. The vertical bars depict the statistical uncertainties.}
\label{fig:promptD}
\end{figure*}

\begin{figure*}[htbp]
\flushleft
\subfigure{
\begin{minipage}[b]{.3\linewidth}
\centering
\begin{overpic}[scale=0.29]{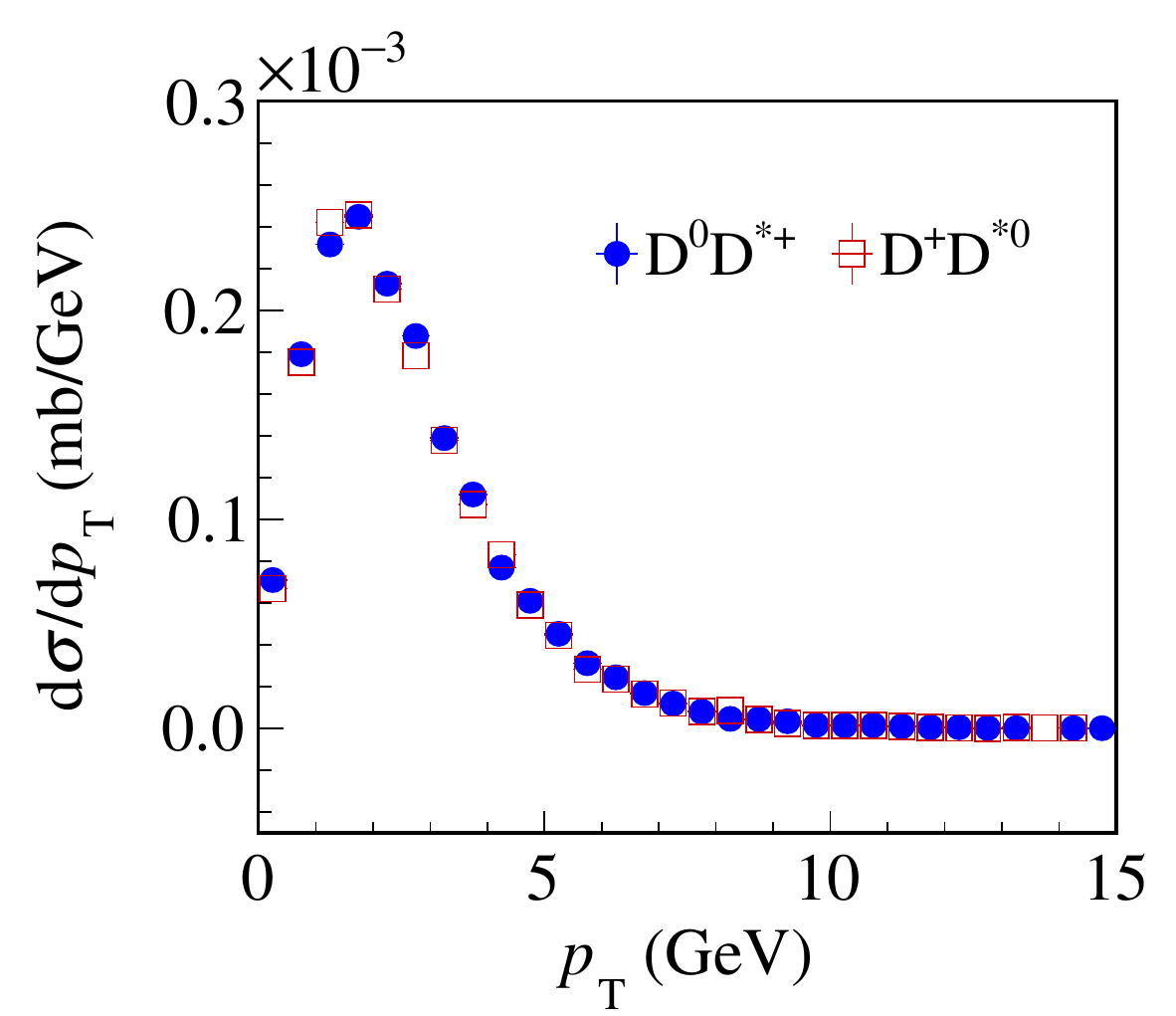}
   \put(3,75){$\textbf{(a)}$}
\end{overpic}
\end{minipage}
}
\subfigure{
\begin{minipage}[b]{.3\linewidth}
\centering
\begin{overpic}[scale=0.29]{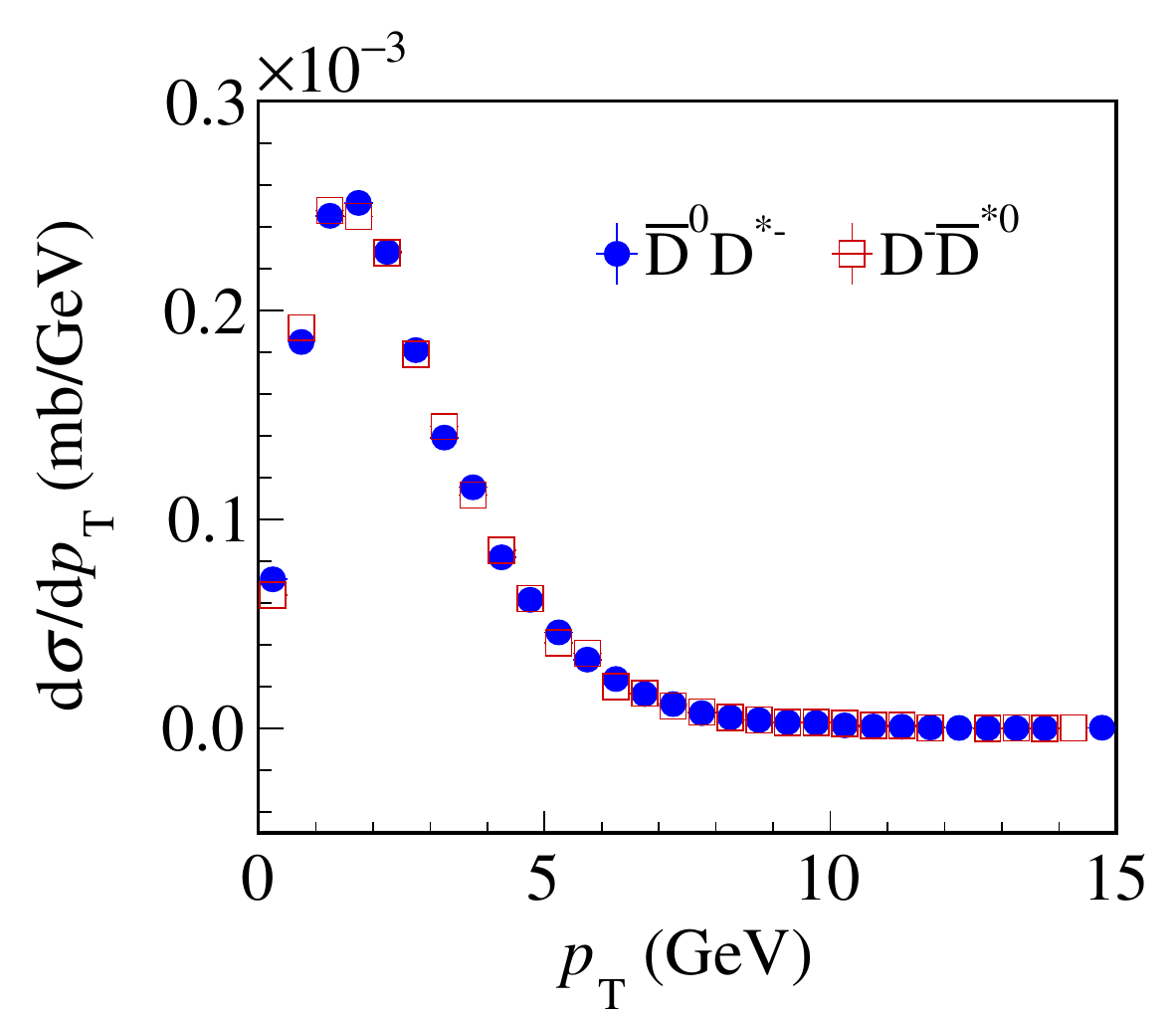}
   \put(3,75){$\textbf{(b)}$}
\end{overpic}
\end{minipage}
}
\subfigure{
\begin{minipage}[b]{.3\linewidth}
\centering
\begin{overpic}[scale=0.29]{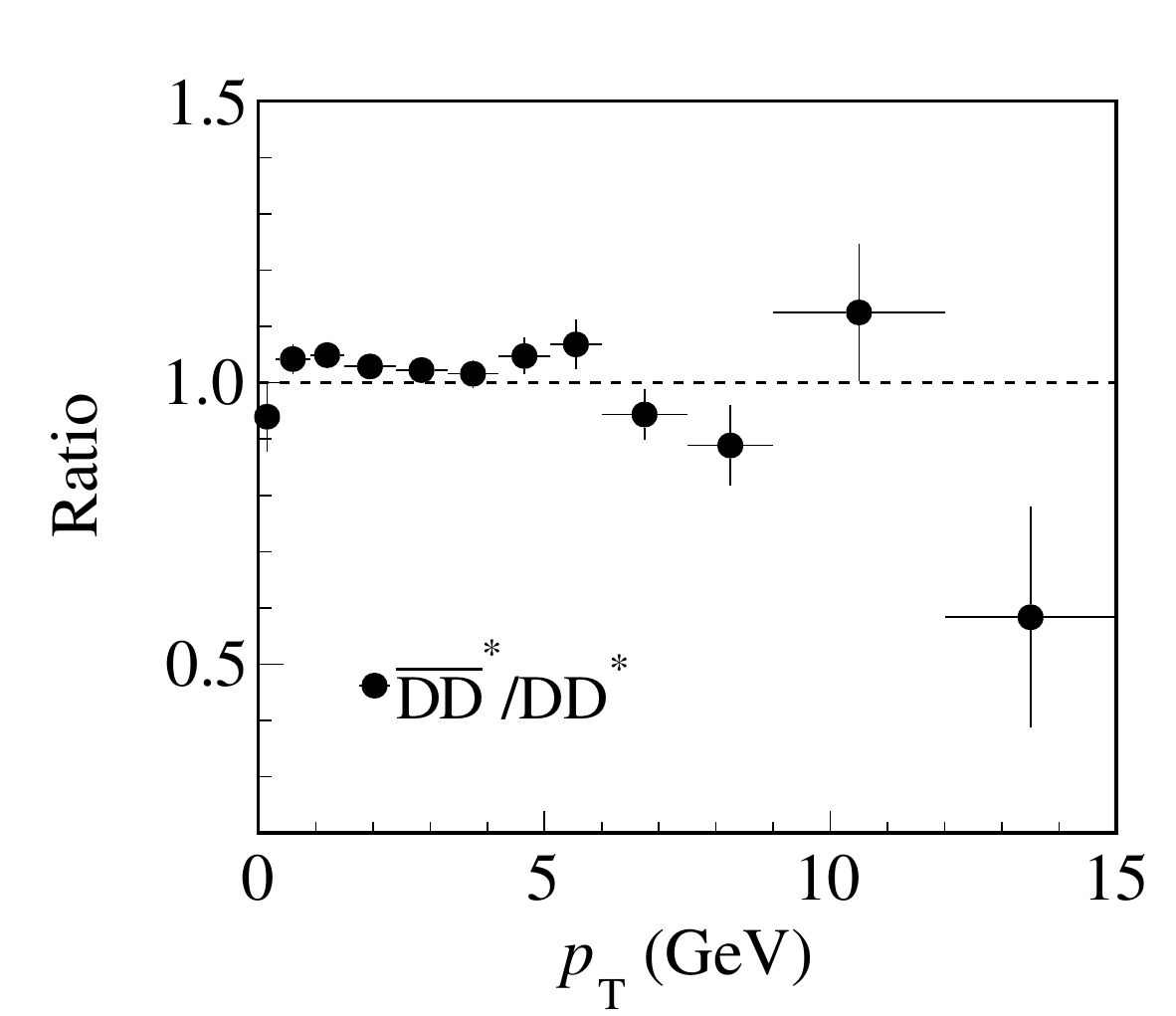}
   \put(3,75){$\textbf{(c)}$}
\end{overpic}
\end{minipage}
}

\subfigure{
\begin{minipage}[b]{.3\linewidth}
\centering
\begin{overpic}[scale=0.29]{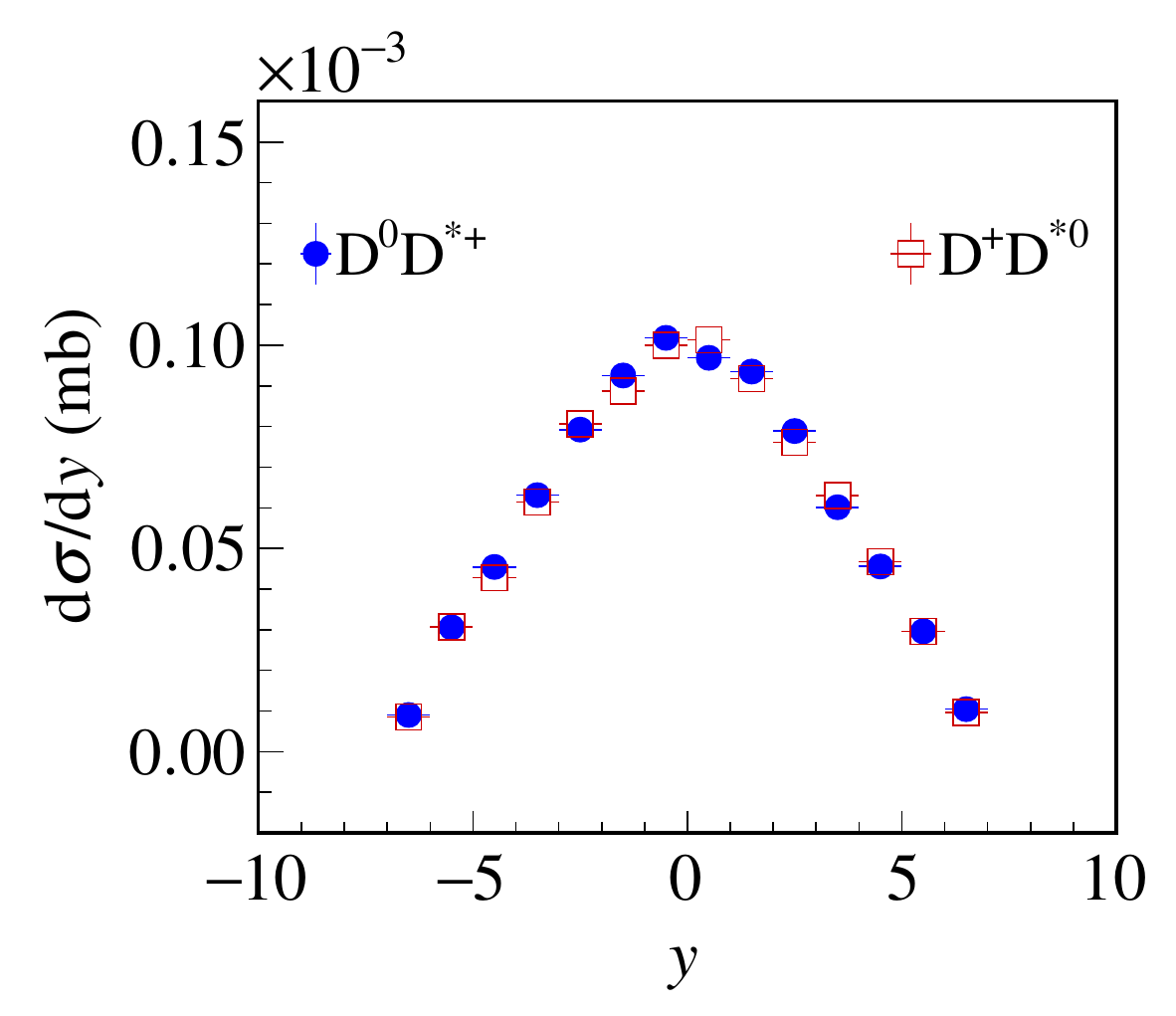}
   \put(3,75){$\textbf{(d)}$}
\end{overpic}
\end{minipage}
}
\subfigure{
\begin{minipage}[b]{.3\linewidth}
\centering
\begin{overpic}[scale=0.29]{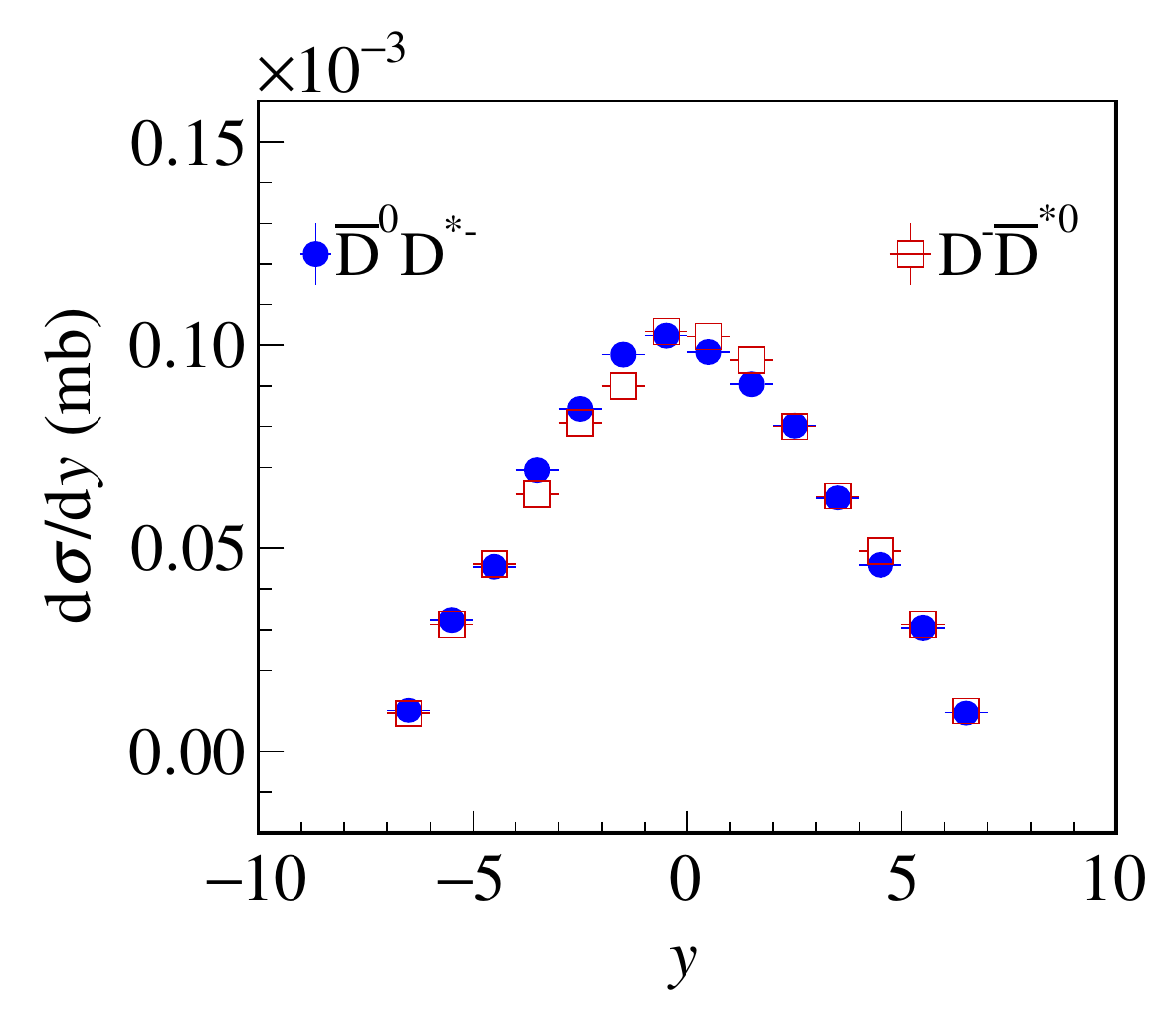}
   \put(3,75){$\textbf{(e)}$}
\end{overpic}
\end{minipage}
}
\subfigure{
\begin{minipage}[b]{.3\linewidth}
\centering
\begin{overpic}[scale=0.29]{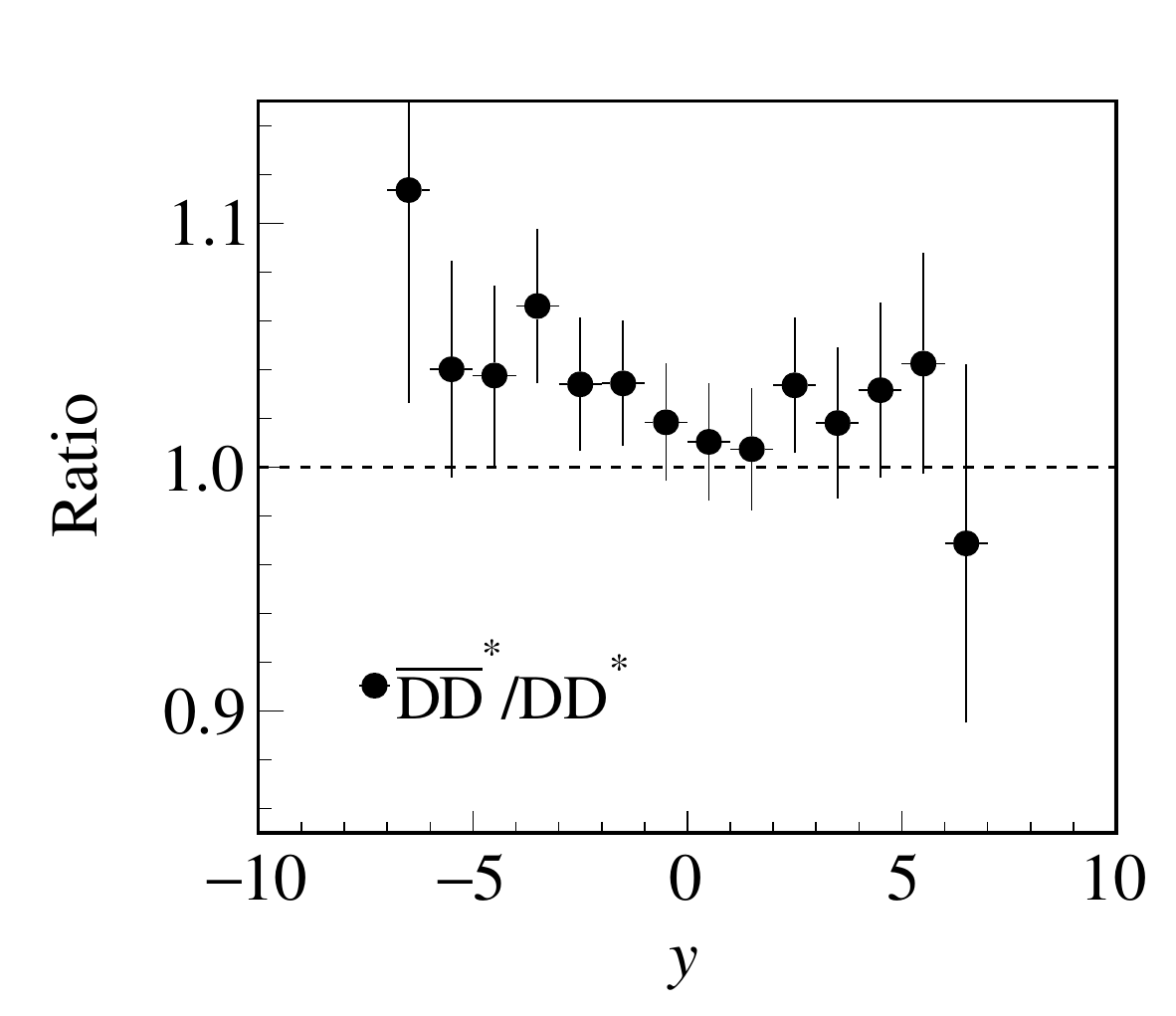}
   \put(3,75){$\textbf{(f)}$}
\end{overpic}
\end{minipage}
}
\caption{The transverse momentum $p_\mathrm{T}$ (a) and the rapidity $y$ (d) distributions of the $D^0D^{\ast +}$ (blue circles) and $D^+D^{*0}$ (red boxes) pairs with relative momentum $k <1~\mathrm{GeV}$ in the $pp$ collisions at $\sqrt{s}=14~\mathrm{TeV}$. Panels (b) and (e) depict the corresponding distributions for their antiparticle pairs. The ratios between the $D^0D^{*+}+D^+D^{*0}$ pair and the $\bar{D}^0D^{*-}+D^-\bar{D}^{*0}$ pair as functions of $p_\mathrm{T}$ and $y$ are shown in panels (c) and (f), respectively. The vertical bars represent the statistical uncertainties.}
\label{fig:DDstarpty}
\end{figure*}

\vspace{0.2cm}
\section{Results and discussions}

Before studying the cross sections of the double charm tetraquark, we first present
the kinematic distributions of the intermediate charmed meson pairs and diquark-antidiquark pairs for the hadronic molecules and the compact tetraquarks in $pp$ collisions at a center-of-mass energy ($\sqrt{s}$) of 14$~\mathrm{TeV}$, respectively. Figure~\ref{fig:promptD}(a) shows the transverse momentum ($p_\mathrm{T}$) distributions of the prompt $D^0$, non-prompt $D^0$ and their combined distribution, while the corresponding distributions for their antiparticles are shown in Fig.~\ref{fig:promptD}(b). One can see that both the distributions decrease significantly when 
$p_\mathrm{T}>2~\mathrm{GeV}$. Accordingly the behavior of the total cross section of single (anti)charmed meson is dominated by that at low $p_\mathrm{T}$ region. 
Figure~\ref{fig:promptD}(c) presents the ratios between the prompt anticharmed meson and the prompt charmed meson, i.e. $\bar{D}^0/D^0$, $D^-/D^+$, $\bar{D}^{\ast 0}/D^{\ast 0}$ and $D^{\ast -}/D^{\ast +}$ as a function of $p_\mathrm{T}$ by blue circles, red squares, purple triangles and orange crosses, respectively. One can see that the ratios are larger 
than one at low $p_\mathrm{T}$ region, 
which is consistent with the expectation from the heavy-quark recombination mechanism~\cite{Braaten:2002yt,Chang:2003ag}.

\begin{figure}[htbp]
\centering
\includegraphics[scale=0.45]{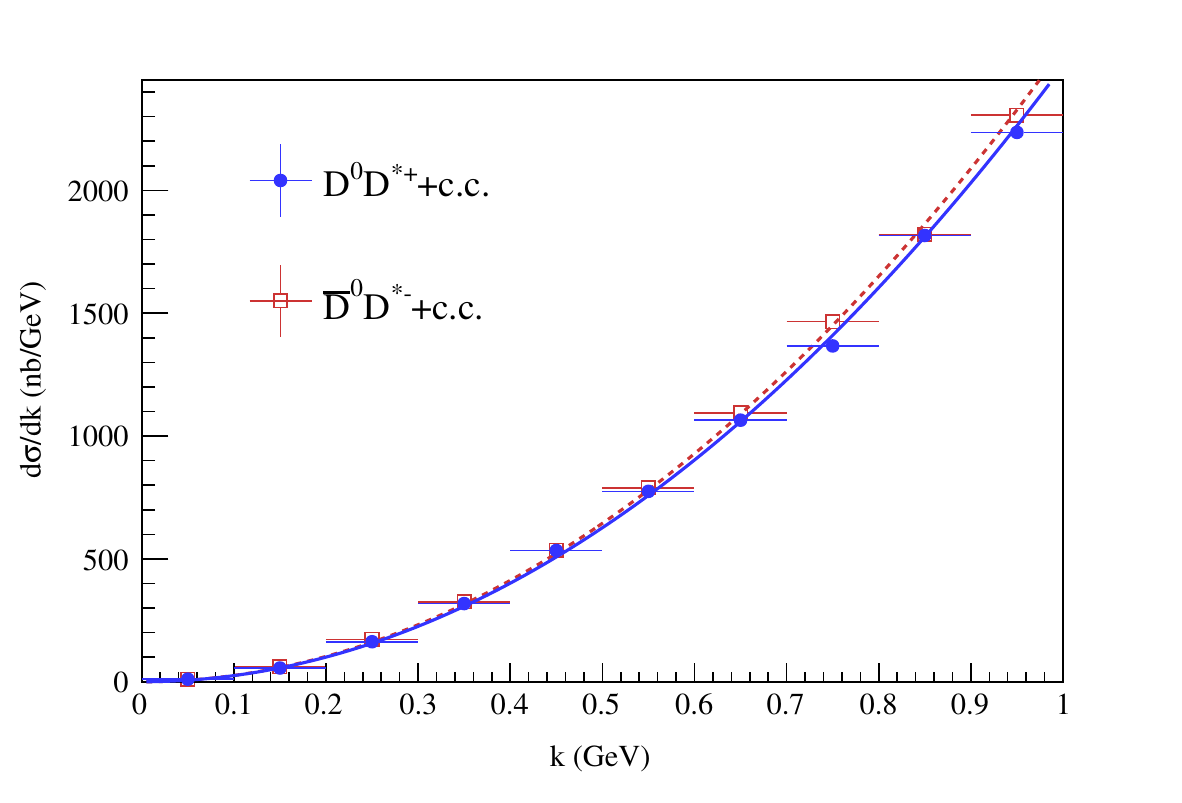}
\caption{The distributions of the relative three-momentum $k$ for the $D^0D^{\ast +}+c.c.$ (blue solid circles) and $\bar{D}^0D^{\ast -}+c.c.$ (red open squares) pairs within the molecular picture are shown. The blue solid and red dashed curves represent the fitting applied to the simulated data. Notably, these two curves exhibit a highly similar dependence of the $k$.}
 \label{fig:kdistribution}
\end{figure}

The $p_\mathrm{T}$ and the rapidity ($y$) distributions of the $D^0D^{\ast +}$ and $D^+D^{\ast 0}$ pairs which form hadronic molecular $T_{cc}^+$ are presented in Figs.~\ref{fig:DDstarpty}(a) and \ref{fig:DDstarpty}(d), respectively. The distributions of their antiparticles are presented in Figs.~\ref{fig:DDstarpty}(b) and \ref{fig:DDstarpty}(e). The distributions of one pair of (anti)charm mesons are similar to each other. Both of them 
decrease when $p_\mathrm{T}$ goes larger than $2~\mathrm{GeV}$. They reach their maximum values at 
$y=0$. 
Meanwhile, the ratios between the $\bar{D}\bar{D}^\ast$ pairs and $DD^\ast$ pairs as functions of $p_\mathrm{T}$ and $y$ are shown in Figs.~\ref{fig:DDstarpty}(c) and \ref{fig:DDstarpty}(f), respectively. In the former figure, 
we can see that the cross section of the anticharmed meson pair is larger than that of the charmed one
at low $p_\mathrm{T}$ region. The case also happens 
for forward region, i.e. large rapidity region. 
On the other hand, since the cross sections of the hadronic molecular $T_{cc}^+$ and $\bar{T}_{\bar{c}\bar{c}}$ are proportional to the differential cross sections $\mathrm{d}\sigma(DD^*)/\mathrm{d}k$ and $\mathrm{d}\sigma(\bar{D}\bar{D}^*)/\mathrm{d}k$ (Eq.\eqref{fomula:6}), respectively, which are presented in Fig.~\ref{fig:kdistribution}. Here the $k$ represents the relative momentum of two charmed hadrons in their rest frame. Although the cross sections of the $DD^*$ and $\bar{D}\bar{D}^*$ pairs display, to a certain extent, asymmetries, their differential cross sections do not exhibit asymmetry.

\begin{figure}[htbp]
\centering
\flushleft
\subfigure{
\begin{minipage}[b]{.9\linewidth}
\includegraphics[scale=0.4]{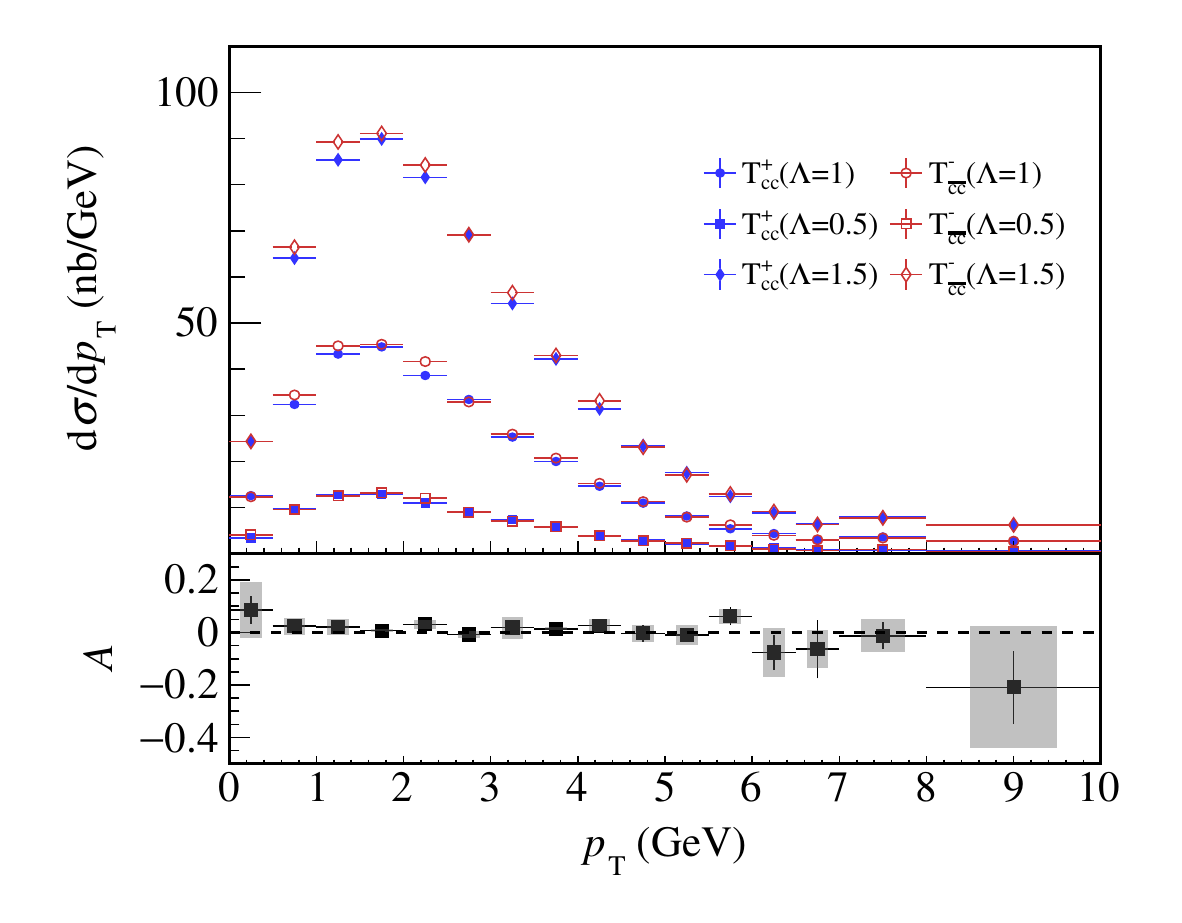}
\end{minipage}
}
\subfigure{
\begin{minipage}[b]{.9\linewidth}
\includegraphics[scale=0.4]{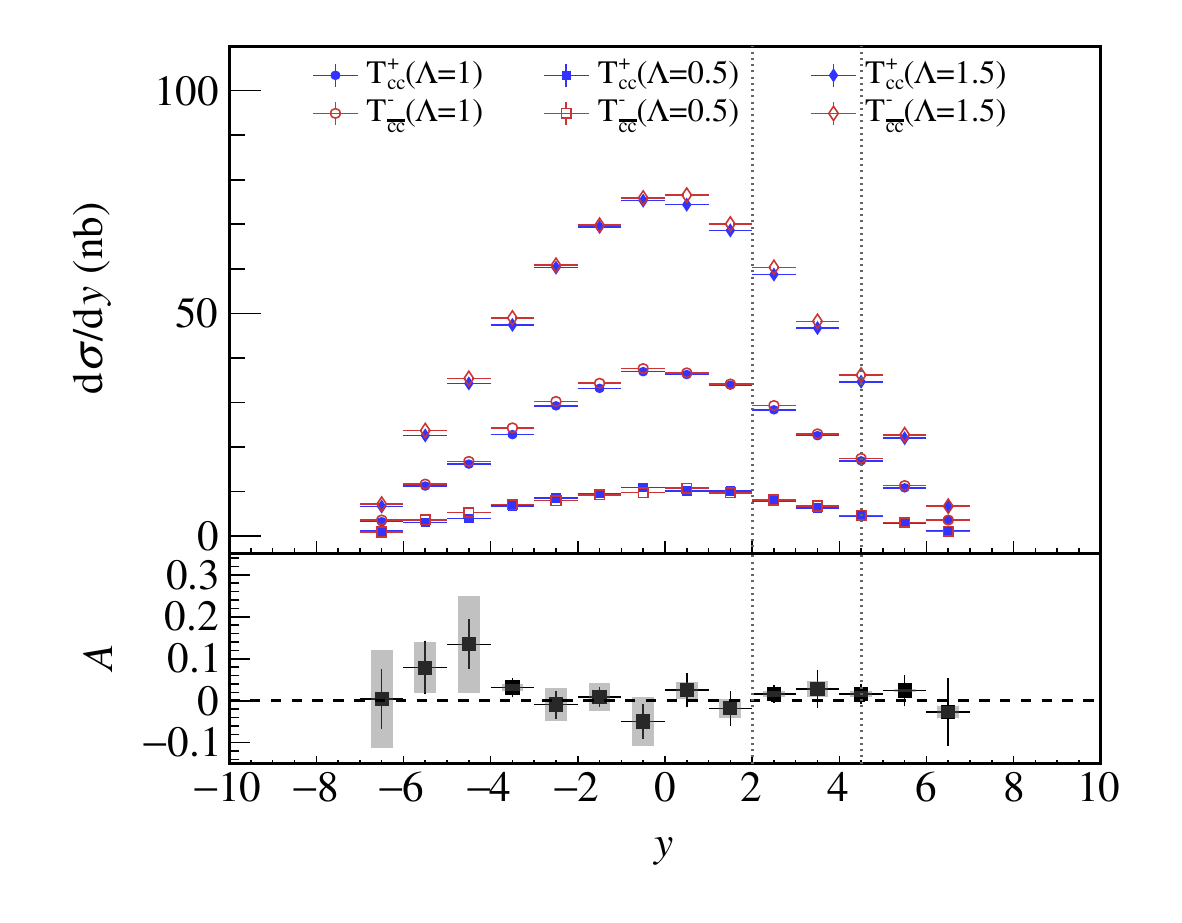}
\end{minipage}
}
\caption{\label{fig:HM}The $p_\mathrm{T}$ (upper) and the $y$ (lower) distributions of the prompt $T_{cc}^{+}$ (blue solid circles) and $T_{\bar{c} \bar{c}}^{-}$ (red open squares) within the molecular framework, in the $pp$ collisions at $\sqrt{s} =$ 14 TeV without any kinematic cuts. The blue (red) solid (open) squares, circles, and diamonds are the distributions of the $T_{cc}^+$ ($T_{\bar{c}\bar{c}}^-$) with 
$\Lambda=0.5~\mathrm{GeV}$, $1~\mathrm{GeV}$ and $1.5~\mathrm{GeV}$, respectively.
The lower panel of each figure represents the asymmetry analogous to that defined in 
Eq.\eqref{eq:A}. The vertical bars and shaded bands are for statistic and systematic uncertainties, respectively.
The two gray vertical dotted lines delineate the acceptance region ($2<y<4.5$) of the LHCb detector.}
\end{figure}

In total, we simulated $1$ billion minimum-bias $pp$ collisions at $\sqrt{s} = 14~\mathrm{TeV}$ with PYTHIA 8.3 for the hadronic molecular picture. The non-diffractive $pp$ cross section at $\sqrt{s} = 14~\mathrm{TeV}$ utilized in this study is $57.17~\mathrm{mb}$, as obtained from PYTHIA. This is also the calibration for the compact tetraquark picture. 
  After filtering the data, the $p_\mathrm{T}$ and $y$ distributions of the $T_{cc}^{+}$ and $T_{\bar{c} \bar{c}}^{-}$ in the molecular picture are presented in Fig.~\ref{fig:HM}.
To estimate the systematic uncertainties, we present the distributions with the three $\Lambda$ values, i.e. $0.5~\mathrm{GeV}$, $1~\mathrm{GeV}$ and $1.5~\mathrm{GeV}$. 
From the figure, one can see that the differential cross section distributions 
of both particles and antiparticles increase with the increasing $\Lambda$, indicating their correlation. To a certain extent, one can see the difference between particle and antiparticle. To see their difference, we introduce a physical quantity, i.e. the asymmetry,
\begin{eqnarray}
     \mathcal{A}\equiv \frac{\sigma^--\sigma^+}{\sigma^-+\sigma^+}
     \label{eq:A}
 \end{eqnarray}
 where the $\sigma^+$ and $\sigma^-$ represent the cross sections of $T_{cc}^+$ and $T_{\bar{c}\bar{c}}^-$, respectively, which 
 have been illustrated by the lower panels of Fig.~\ref{fig:HM}. The central value is defined as the weighted average of the asymmetry of three different choices of $\Lambda$
 \begin{equation}
\mathcal{A}\equiv \omega_1\mathcal{A}_1+\omega_2\mathcal{A}_2+\omega_3\mathcal{A}_3,
\label{eq:A}
 \end{equation}
 with 
 \begin{equation}
\omega_i=\frac{\frac{1}{\delta_i^2}}{\frac{1}{\delta_1^2}+\frac{1}{\delta_2^2}+\frac{1}{\delta_3^2}},
 \end{equation}
 where $\delta_i$ represents the relative statistic error of the asymmetry with the $i$th choice. 
The statistic error of the asymmetry is also defined as the weighted average of 
the three statistic errors with three different choices of $\Lambda$
 \begin{equation}
\delta_{sta}\equiv \omega_1\delta_1+\omega_2\delta_2+\omega_3\delta_3.
 \end{equation}
 The systematic error is defined as the root mean square value of
 the three deviations of the three different $\mathcal{A}_i$ from the central value defined in Eq,\eqref{eq:A}.
  \begin{equation}
\delta_{sys}\equiv \sqrt{\frac{\sum_i (\mathcal{A}_i-\mathcal{A})^2}{3}}.
 \end{equation}
The similarity of particles and antiparticles  
 is because a $c$ quark ($\bar{c}$ quark) initially combines with a light antiquark (quark) to create a $D^{(*)}$ ($\bar{D}^{(*)}$) meson, and subsequently, two of these mesons combine to yield a $T_{cc}^+$ ($T_{\bar{c} \bar{c}}^-$). As a result, the phase space of the 
$D$ and $D^*$ meson needs to be rigorously considered,
making the cross section of $T_{cc}$ being proportional to $\mathrm{d}\sigma(DD^*)/\mathrm{d}k$~\cite{Guo:2014qra,Guo:2013ufa}, as shown in Eq.~\eqref{fomula:6}, instead of $\sigma(DD^*)$. Although the $\sigma(\bar{D}\bar{D}^*)$ and $\sigma(DD^*)$ have significantly different behaviors, which stems from the different behaviors of the single anticharmed meson and charmed meson, 
their dependence on the relative momentum are almost the same.
 They reach their maximum values at $p_\mathrm{T}\approx 2~\mathrm{GeV}$,
and then decrease dramatically. Their $y$ distributions are symmetric and reach 
their maximum values at $y \approx 0$. 
The systematic uncertainties in the molecular picture are estimated by altering $\Lambda$ from $0.5~\mathrm{GeV}$ to $1.5~\mathrm{GeV}$ in Eq.~\eqref{fomula:2}, which are only used to illustrate the correlation of the systematic uncertainties in the asymmetry defined in the following.
 It is noteworthy that those uncertainties for particle and antiparticle are strongly correlated with each other. 

\begin{figure*}[htbp]
\flushleft
\subfigure{
\begin{minipage}[b]{.3\linewidth}
  \centering
  \begin{overpic}[scale=0.29]{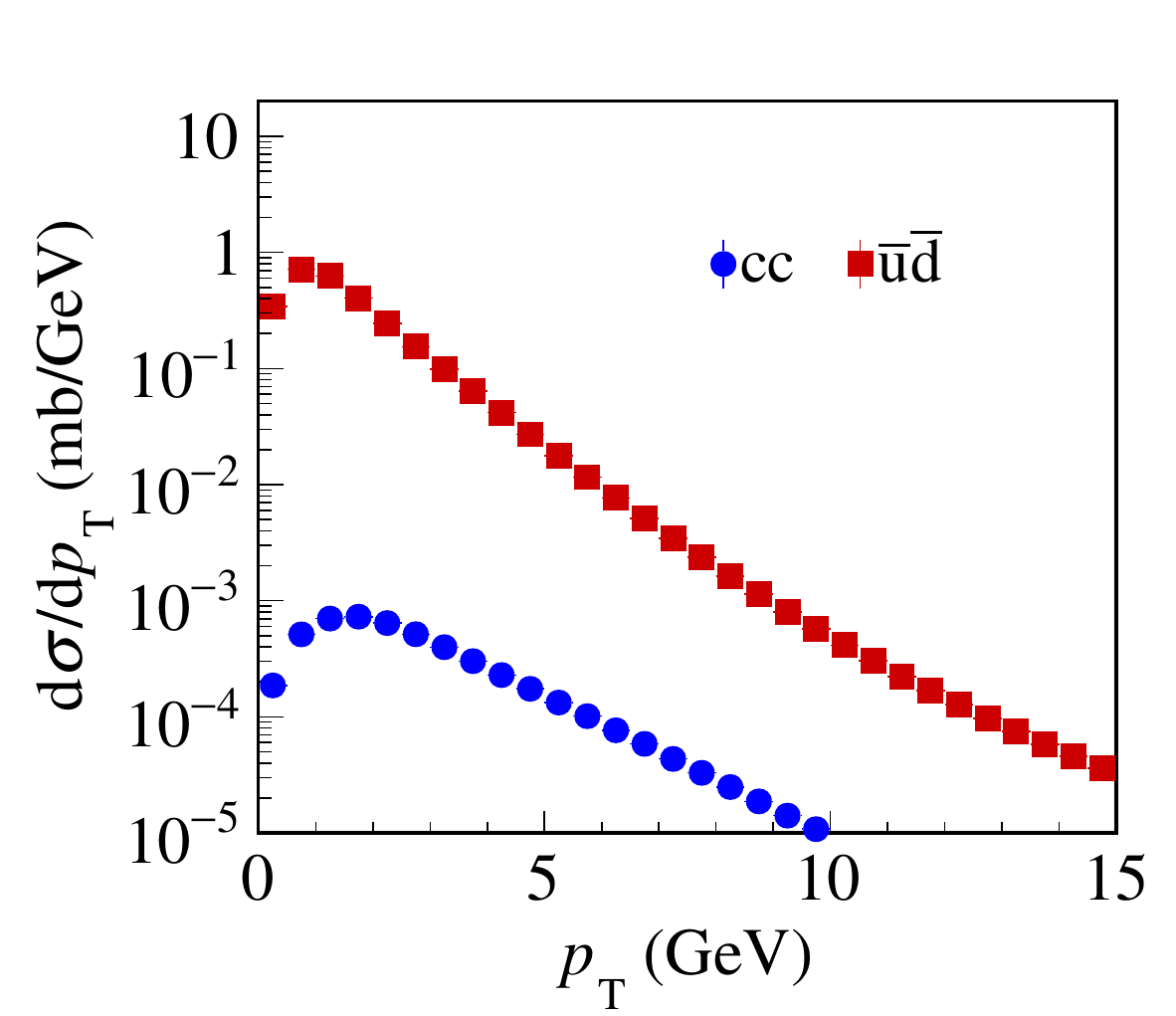}
   \put(2.5,75){$\textbf{(a)}$}
  \end{overpic}
\end{minipage}
}
\subfigure{
\begin{minipage}[b]{.3\linewidth}
  \centering
  \begin{overpic}[scale=0.29]{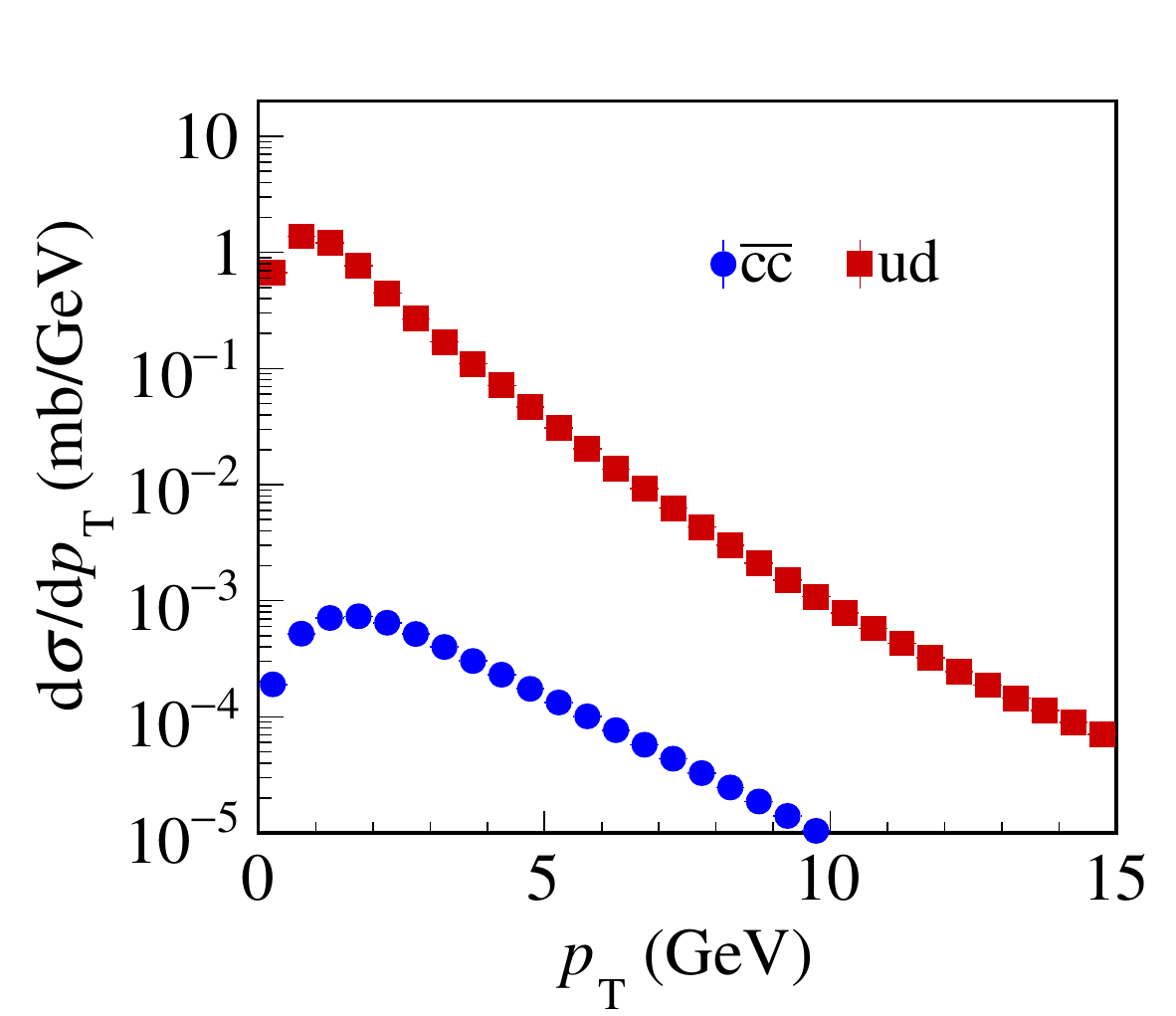}
   \put(2.5,75){$\textbf{(b)}$}
  \end{overpic}
\end{minipage}
}
\subfigure{
\begin{minipage}[b]{.3\linewidth}
  \centering
  \begin{overpic}[scale=0.305]{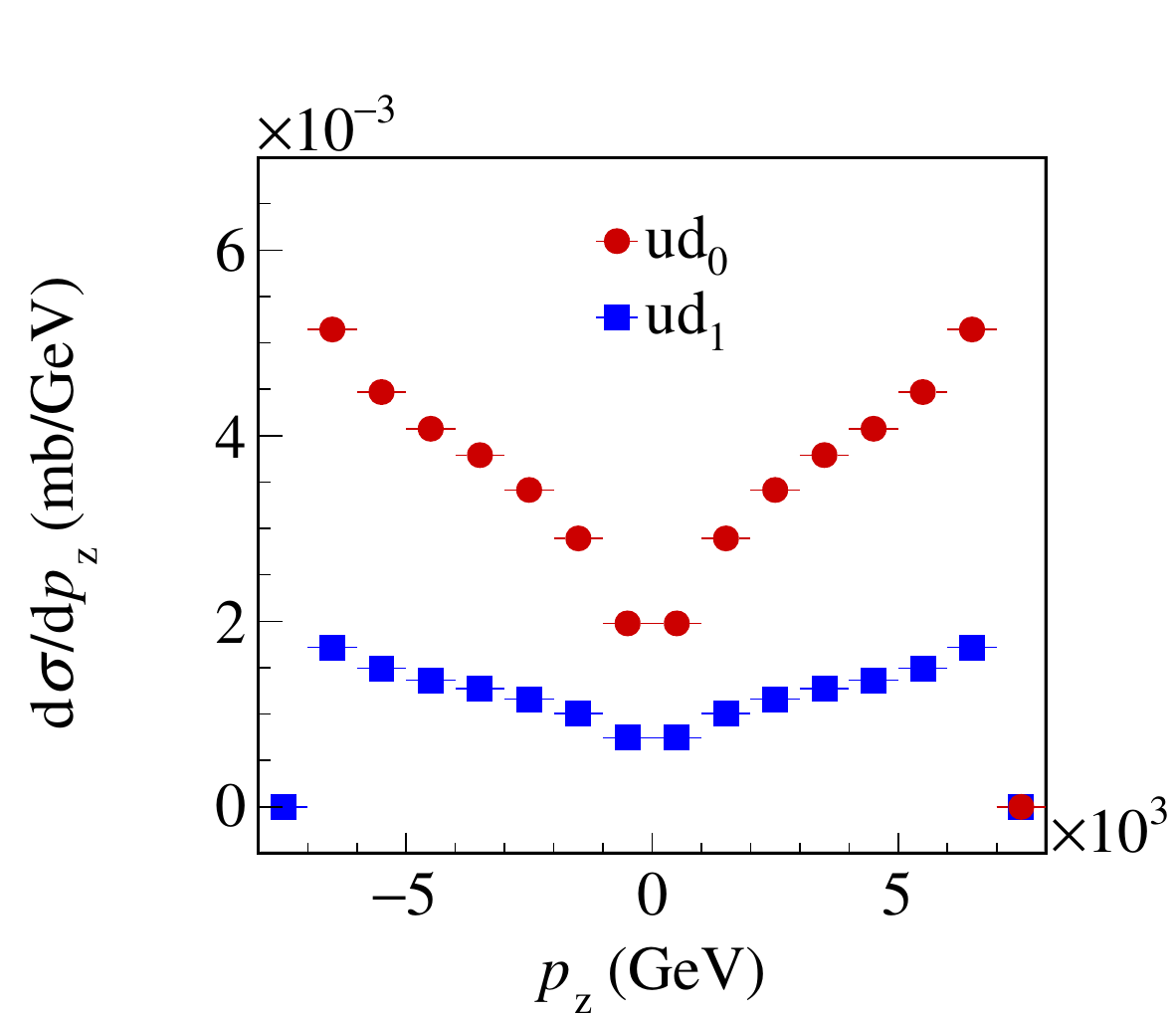}
   \put(2.5,71){$\textbf{(c)}$}
  \end{overpic}
\end{minipage}
}

\subfigure{
\begin{minipage}[b]{.3\linewidth}
  \centering
  \begin{overpic}[scale=0.29]{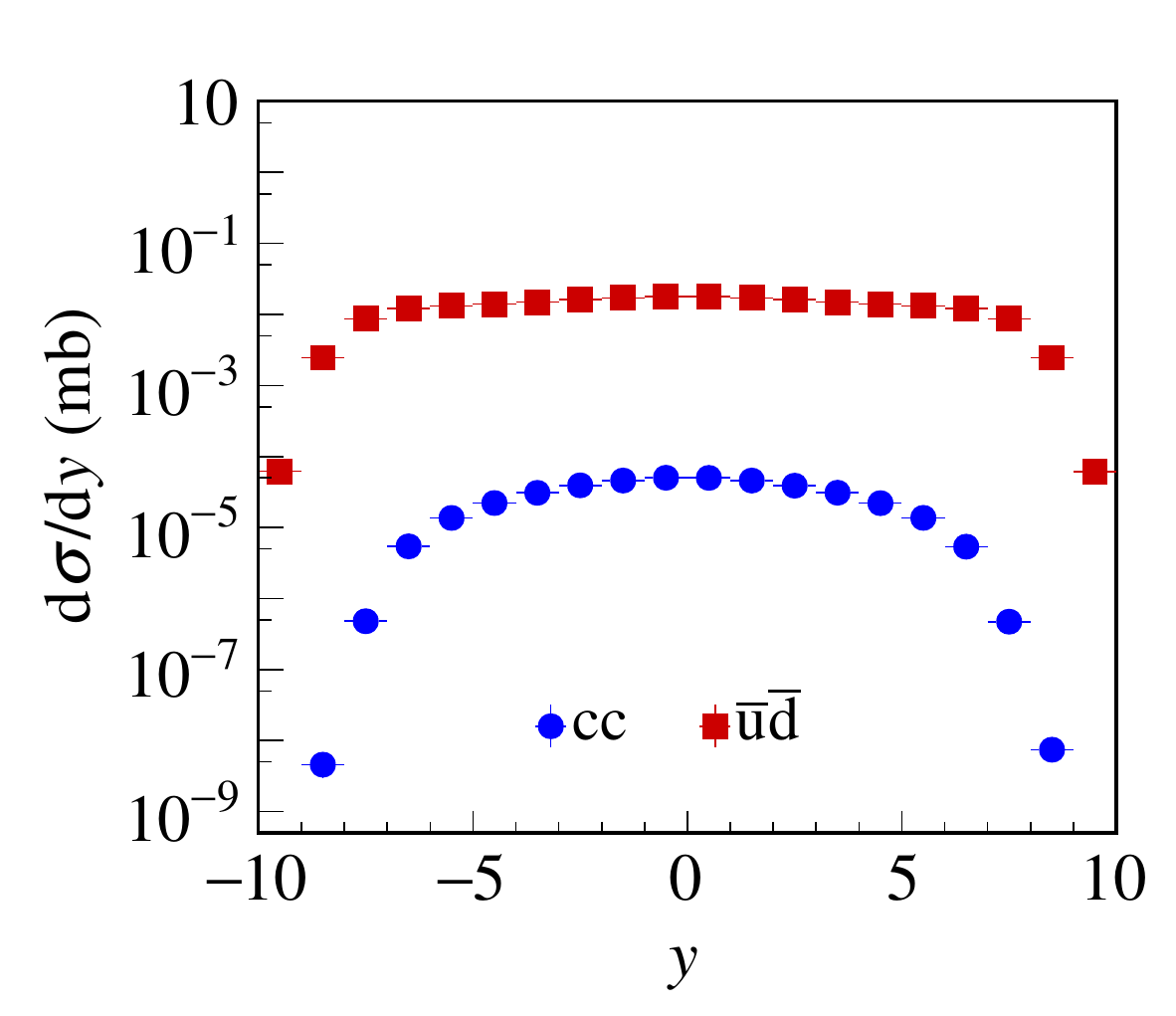}
   \put(2.5,75){$\textbf{(d)}$}
  \end{overpic}
\end{minipage}
}
\subfigure{
\begin{minipage}[b]{.3\linewidth}
  \centering
  \begin{overpic}[scale=0.29]{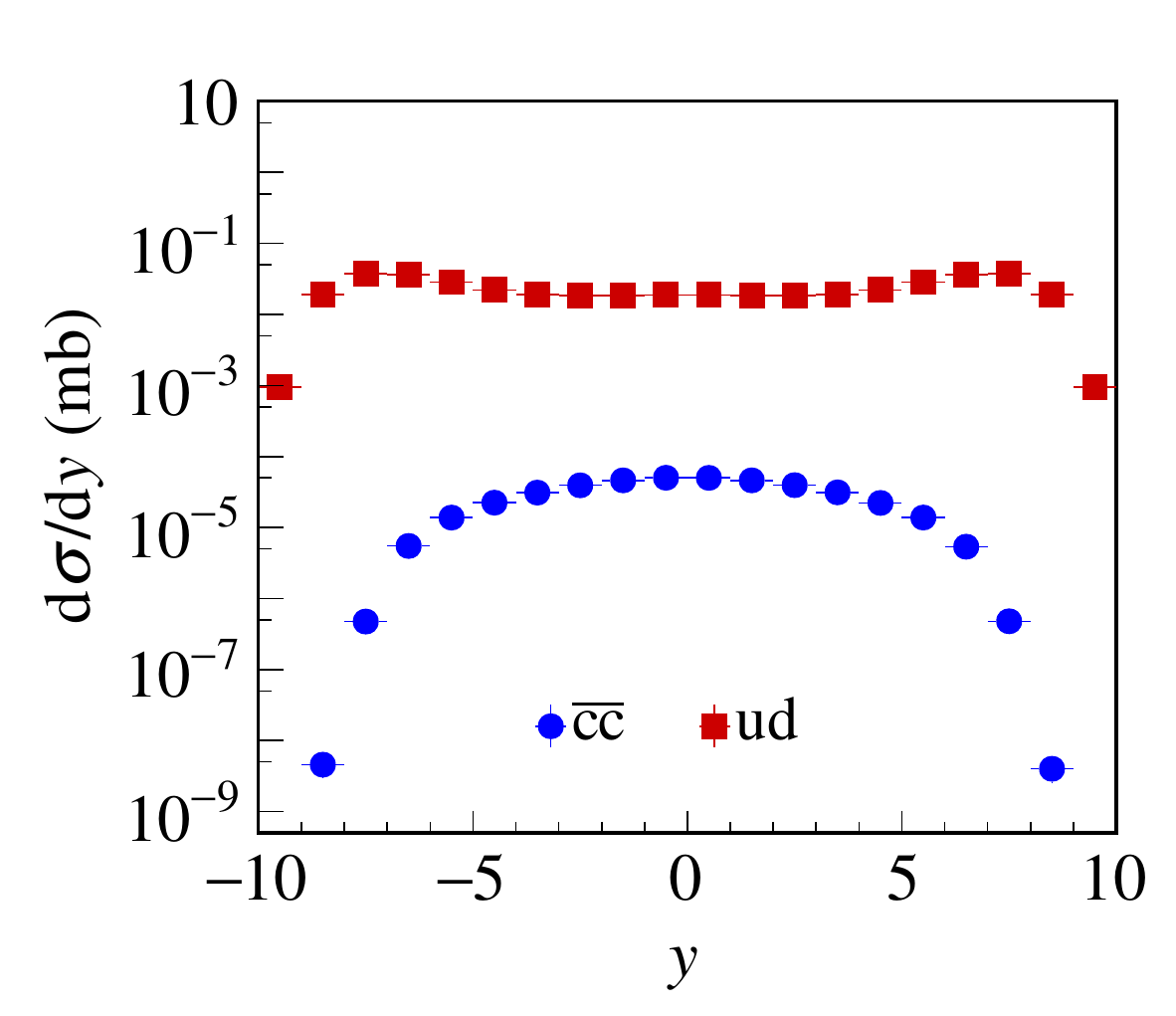}
   \put(2.5,75){$\textbf{(e)}$}
  \end{overpic}
\end{minipage}
}
\subfigure{
\begin{minipage}[b]{.3\linewidth}
  \centering
  \begin{overpic}[scale=0.305]{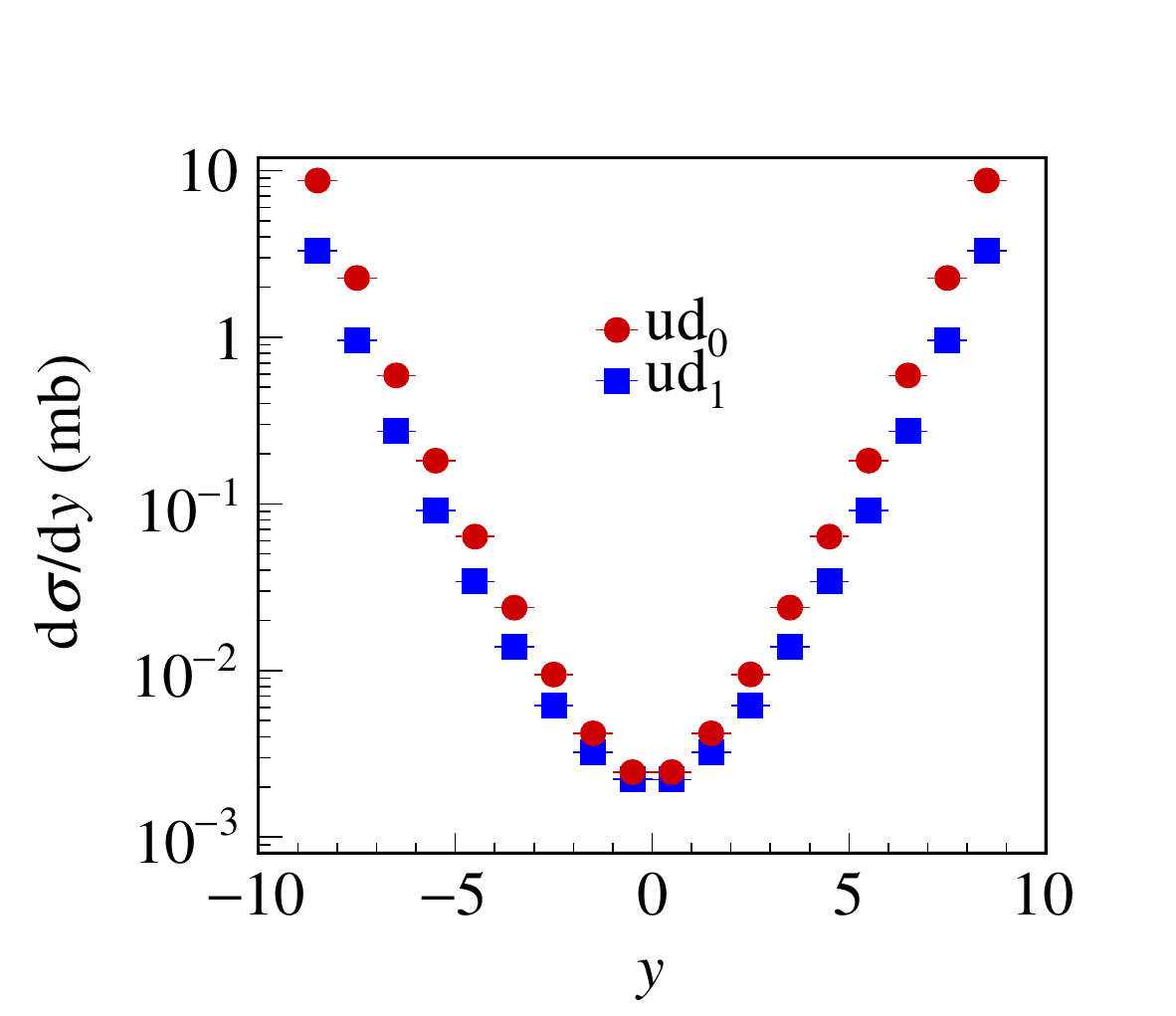}
   \put(2.5,71){$\textbf{(f)}$}
  \end{overpic}
\end{minipage}
}
\caption{Panels (a) and (d) represent the $p_T$ (a) and $y$ (d) distributions for $cc$ (blue circles) and $\bar{u}\bar{d}$ (red squares) diquarks with the relative momentum between $c$ ($\bar{u}$) and $c$ ($\bar{d}$) quarks less than $1~\mathrm{GeV}$ in their rest frame, respectively. The corresponding distributions for $\bar{c}\bar{c}$ (blue circles) and $ud$ (red squares) diquarks are presented in panels (b) and (e). The $p_z$ (c) and $y$ (f) distributions of the diquarks $ud_0$ (red circles) and $ud_1$ (blue squares), with the subscripts $0$ and $1$ indicating the total spin of the diquarks, are obtained directly from Monte Carlo simulation.}
\label{fig:Qptpzy}
\end{figure*}

\begin{figure*}[htbp]
\flushleft
\subfigure{
\begin{minipage}[b]{.3\linewidth}
\centering
\begin{overpic}[scale=0.29]{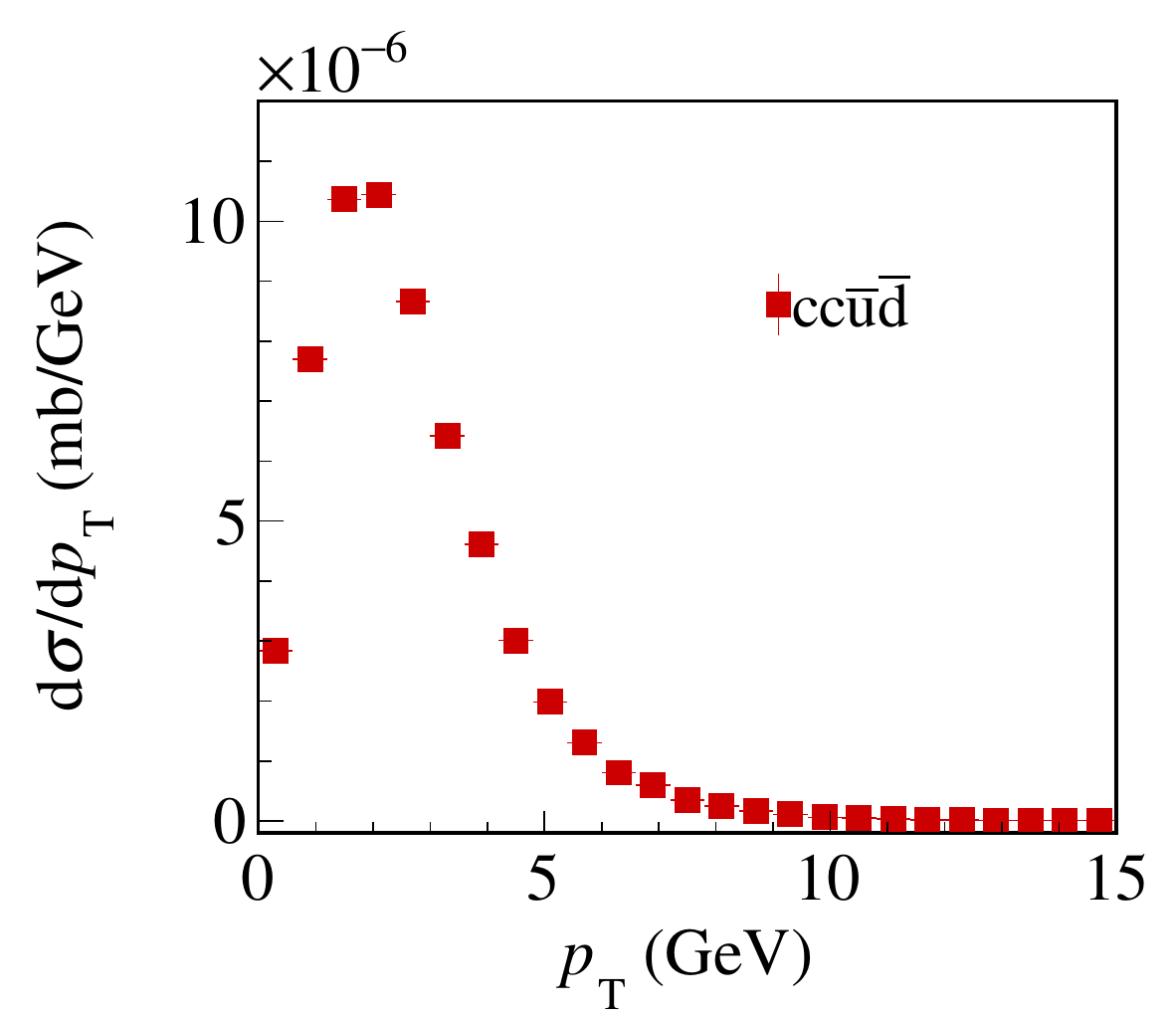}
   \put(3,75){$\textbf{(a)}$}
\end{overpic}
\end{minipage}
}
\subfigure{
\begin{minipage}[b]{.3\linewidth}
\centering
\begin{overpic}[scale=0.29]{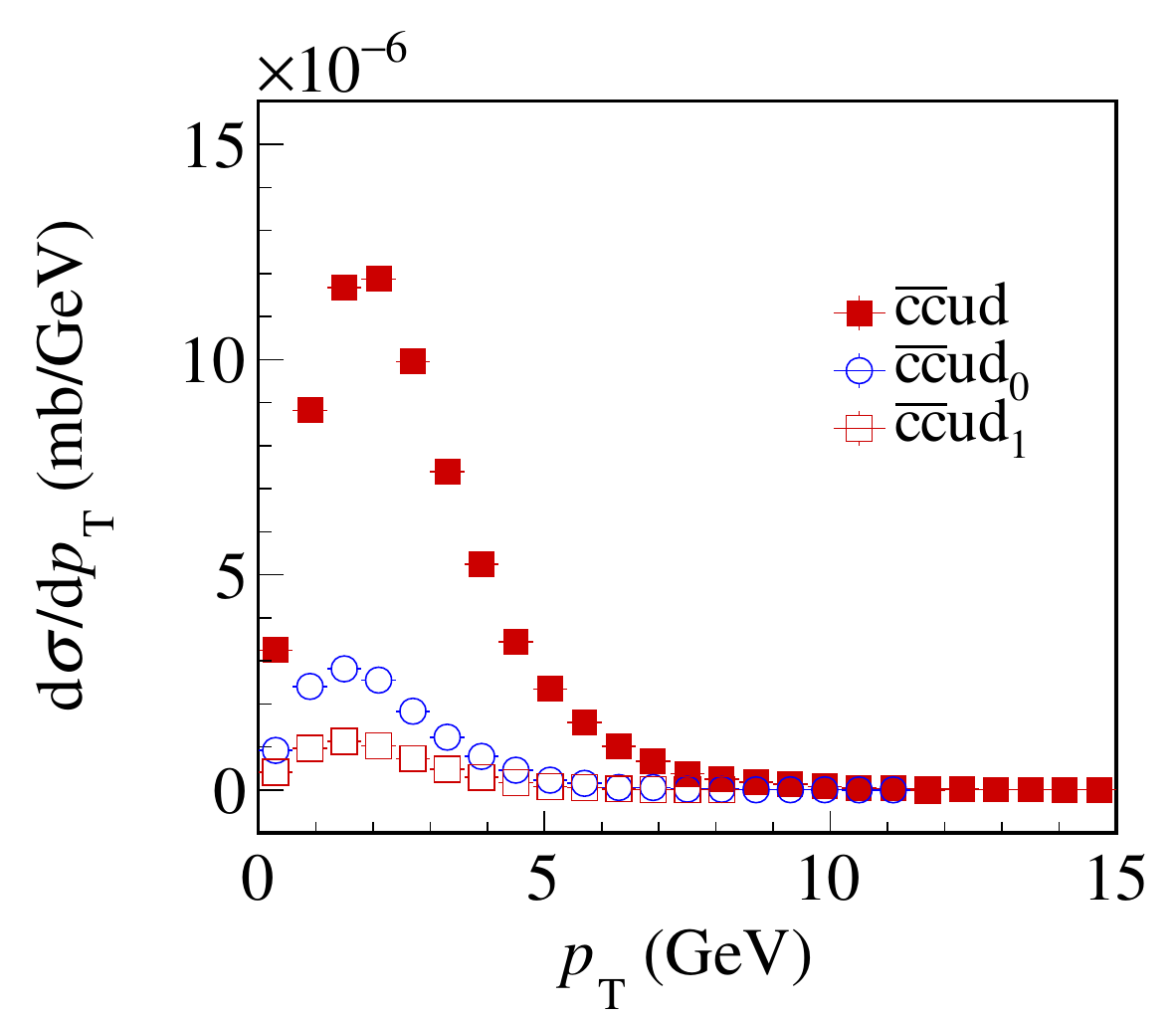}
   \put(3,75){$\textbf{(b)}$}
\end{overpic}
\end{minipage}
}
\subfigure{
\begin{minipage}[b]{.3\linewidth}
\centering
\begin{overpic}[scale=0.29]{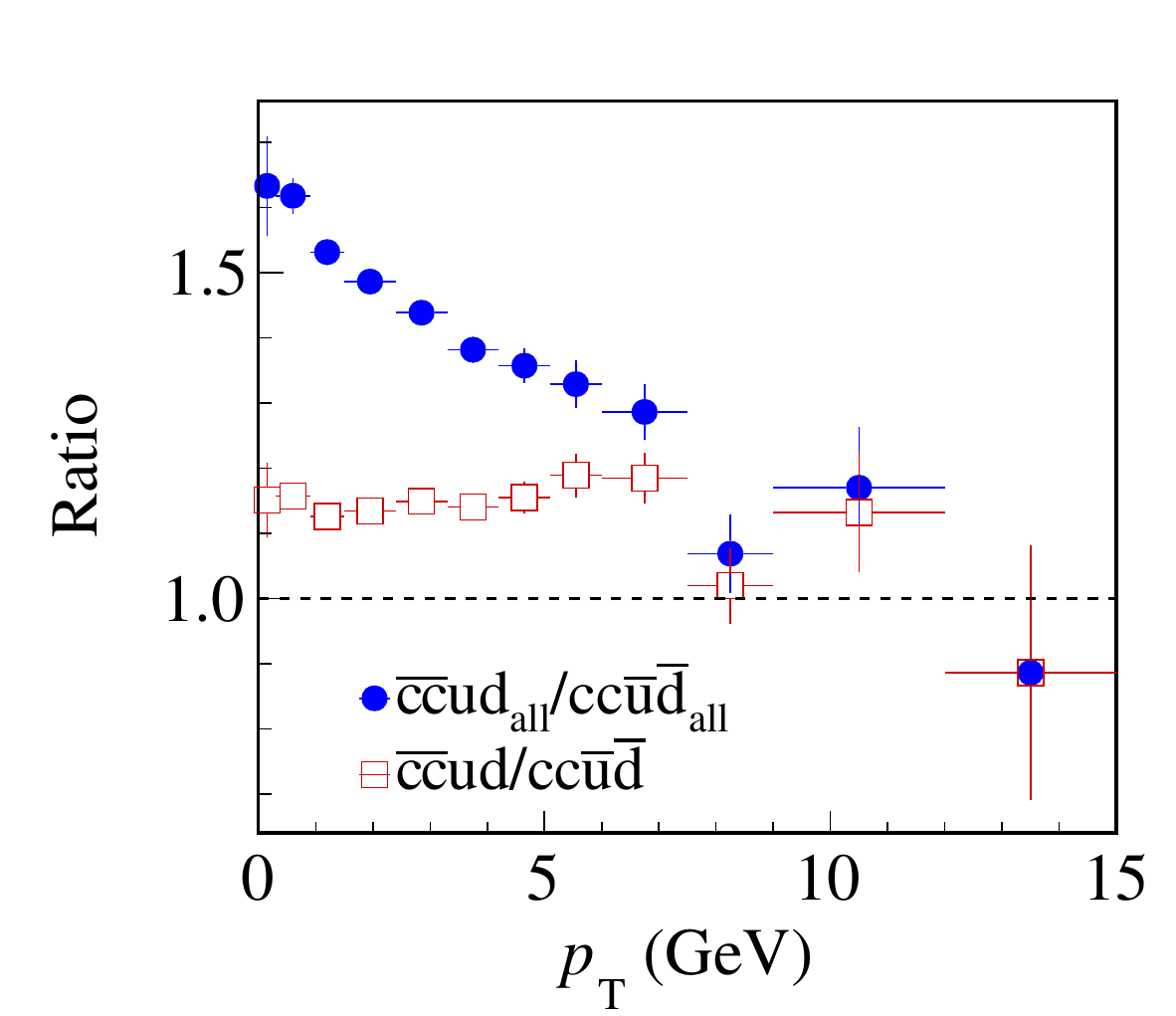}
   \put(3,75){$\textbf{(c)}$}
\end{overpic}
\end{minipage}
}

\subfigure{
\begin{minipage}[b]{.3\linewidth}
\centering
\begin{overpic}[scale=0.29]{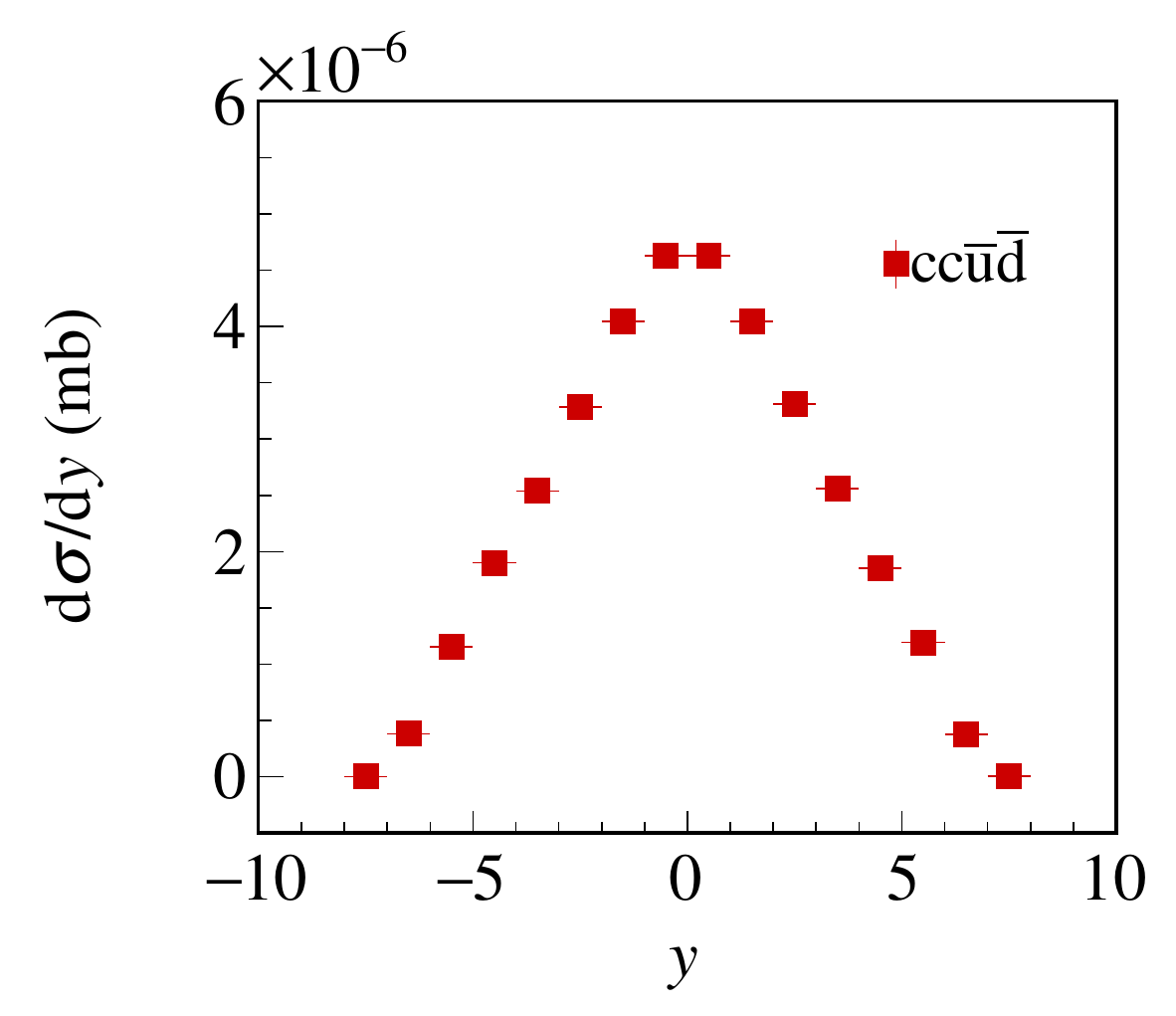}
   \put(3,75){$\textbf{(d)}$}
\end{overpic}
\end{minipage}
}
\subfigure{
\begin{minipage}[b]{.3\linewidth}
\centering
\begin{overpic}[scale=0.29]{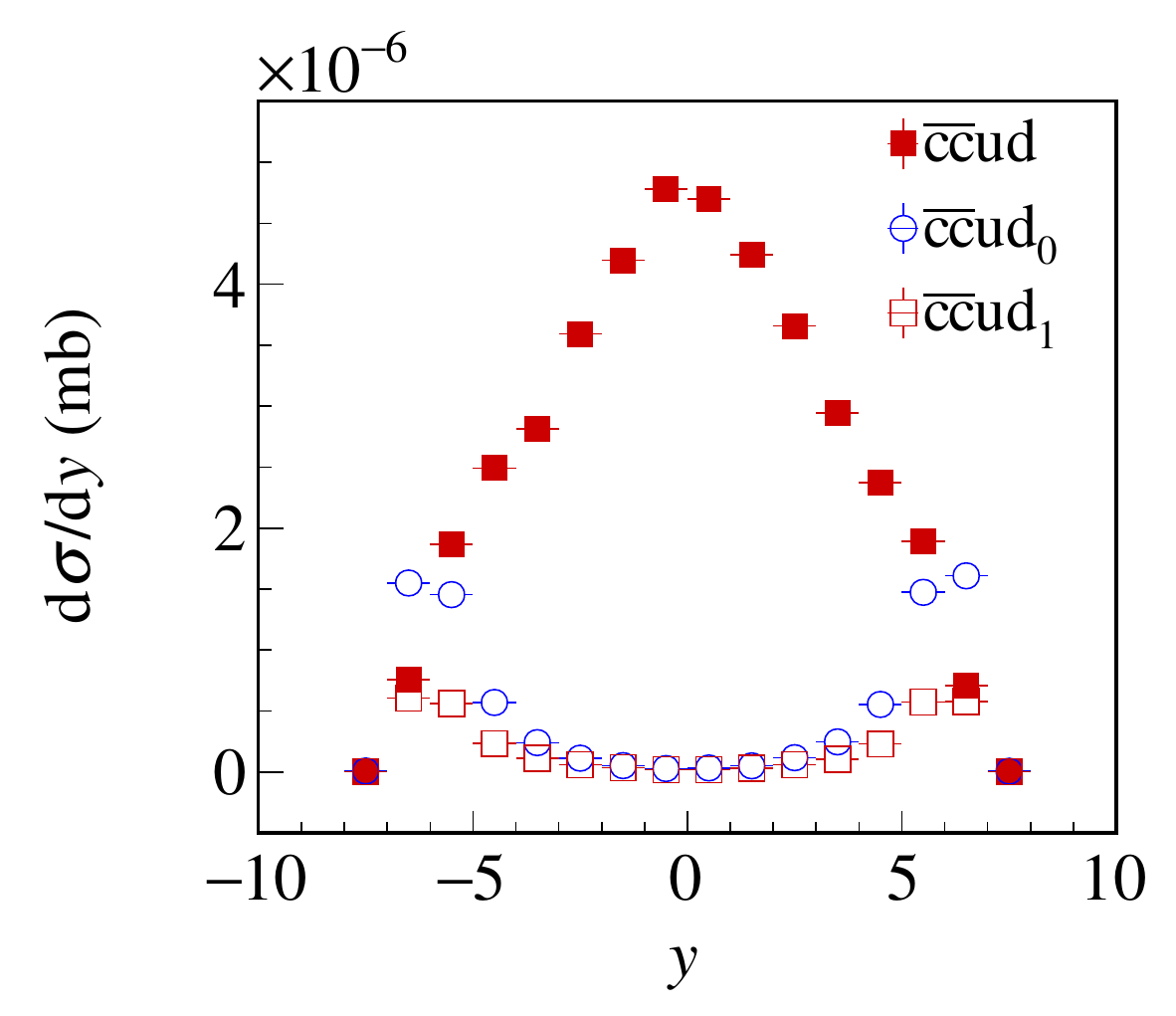}
   \put(3,75){$\textbf{(e)}$}
\end{overpic}
\end{minipage}
}
\subfigure{
\begin{minipage}[b]{.3\linewidth}
\centering
\begin{overpic}[scale=0.29]{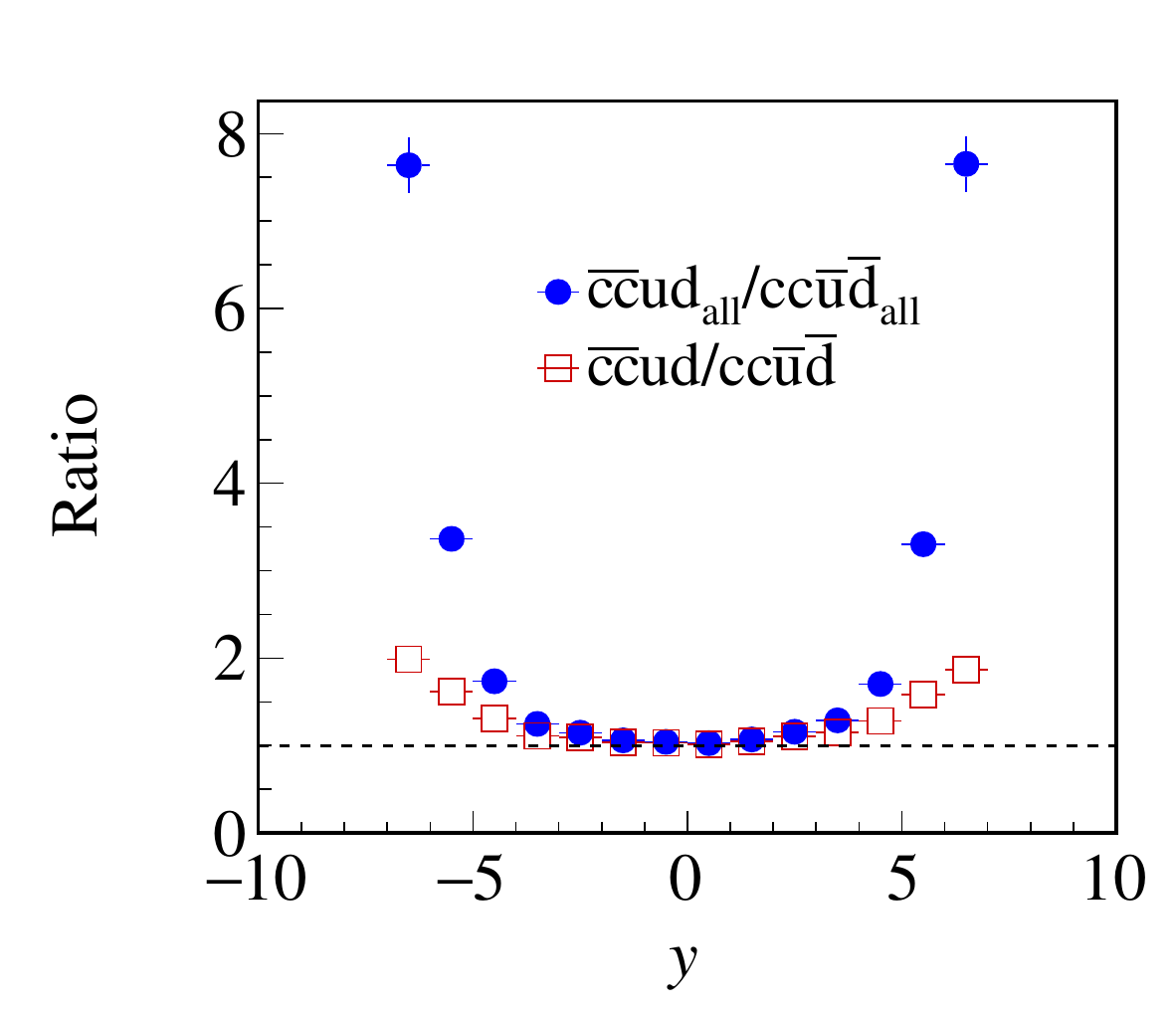}
   \put(3,75){$\textbf{(f)}$}
\end{overpic}
\end{minipage}
}
\caption{The $p_\mathrm{T}$ (a) and $y$ (d) distributions of the charmed diquark and light antidiquark pairs with $k<1~\mathrm{GeV}$ in the $pp$ collisions at $\sqrt{s} = 14~\mathrm{TeV}$. Similar distributions for charmed antidiquark and light diquark pairs are shown in panels (b) and (e), respectively. The distributions for light diquarks $ud_0$ and $ud_1$ directly produced from Monte Carlo simulation, along with $ud$ selected with $k<1~\mathrm{GeV}$, are represented by hollow blue circles, hollow red squares, and solid red squares, respectively. Ratios between $\bar{c}\bar{c}ud$ and $cc\bar{u}\bar{d}$, where diquarks and antidiquarks are manually collected, as functions of $p_\mathrm{T}$ and $y$ are presented with red hollow squares in panels (c) and (f), respectively. Those including light diquarks from Monte Carlo simulation are shown by solid blue circles. The vertical bars depict the statistical uncertainties.}
\label{fig:tccpty}
\end{figure*}

 For the compact tetraquark picture, 
the $p_\mathrm{T}$ and $y$ distributions of the charm diquark and light antidiquark are, in advance, presented in Figs.~\ref{fig:Qptpzy}(a) and \ref{fig:Qptpzy}(d), respectively. They present a similar distribution pattern.
The yield of light antidiquarks $\bar{u}\bar{d}$ is three order-of-magnitude larger than that of charm diquarks. 
In this case, the charm diquarks and light antidiquarks are selected by imposing a condition on the relative momentum between two (anti)quarks of diquark (antidiquark) less than $1~\mathrm{GeV}$ (Tab.~\ref{table:cs-CT1} and, Tab.~\ref{table:cs-CT2}) and $0.5~\mathrm{GeV}$ (Tab.~\ref{table:cs-CT3} and, Tab.~\ref{table:cs-CT4}) in their rest frame for a analysis of systematic uncertainty. The corresponding distributions for their antiparticles are shown in Figs.~\ref{fig:Qptpzy}(b) and \ref{fig:Qptpzy}(e). One can see significant enhancement of the $ud$ light diquark at large rapidity region. 
Additionally, the longitudinal momentum $p_z$ and $y$ distributions for the $ud_0$ and $ud_1$ diquarks directly from Monte Carlo simulation are displayed in Figs.~\ref{fig:Qptpzy}(c) and \ref{fig:Qptpzy}(f), respectively. Those light diquarks from Monte Carlo simulation exhibit very noticeable enhancements at large rapidity region. It is very nature, as the non-interaction remaining light diquarks fly along the beam direction. 
Utilizing these diquarks and antidiquarks, their pair distributions are presented in Fig.~\ref{fig:tccpty}. The distributions of the $cc\bar{u}\bar{d}$ charmed diquark and light antidiquark pairs is the same as that of normal hadrons (Fig.~\ref{fig:tccpty}(a) and (d)). Those of the $\bar{c}\bar{c}ud$ anticharmed diquark and light diquark pairs
are different. Besides the maximum value at $y=0$, there are two significant enhancements around $y=\pm 6$. The ratio of the yield of $\bar{c}\bar{c}ud$ and $cc\bar{u}\bar{d}$ are presented in Fig.~\ref{fig:tccpty}(c) and (f). One can see that the ratios are much larger than one, even without considering those light diquarks from Monte Carlo simulation (hollow red squares). 
One notice that the central value of last point in (c) is smaller than one, but with large statistic uncertainty. 

\begin{figure}[htbp]
\flushleft
\subfigure{
\begin{minipage}[b]{.9\linewidth}
\includegraphics[scale=0.4]{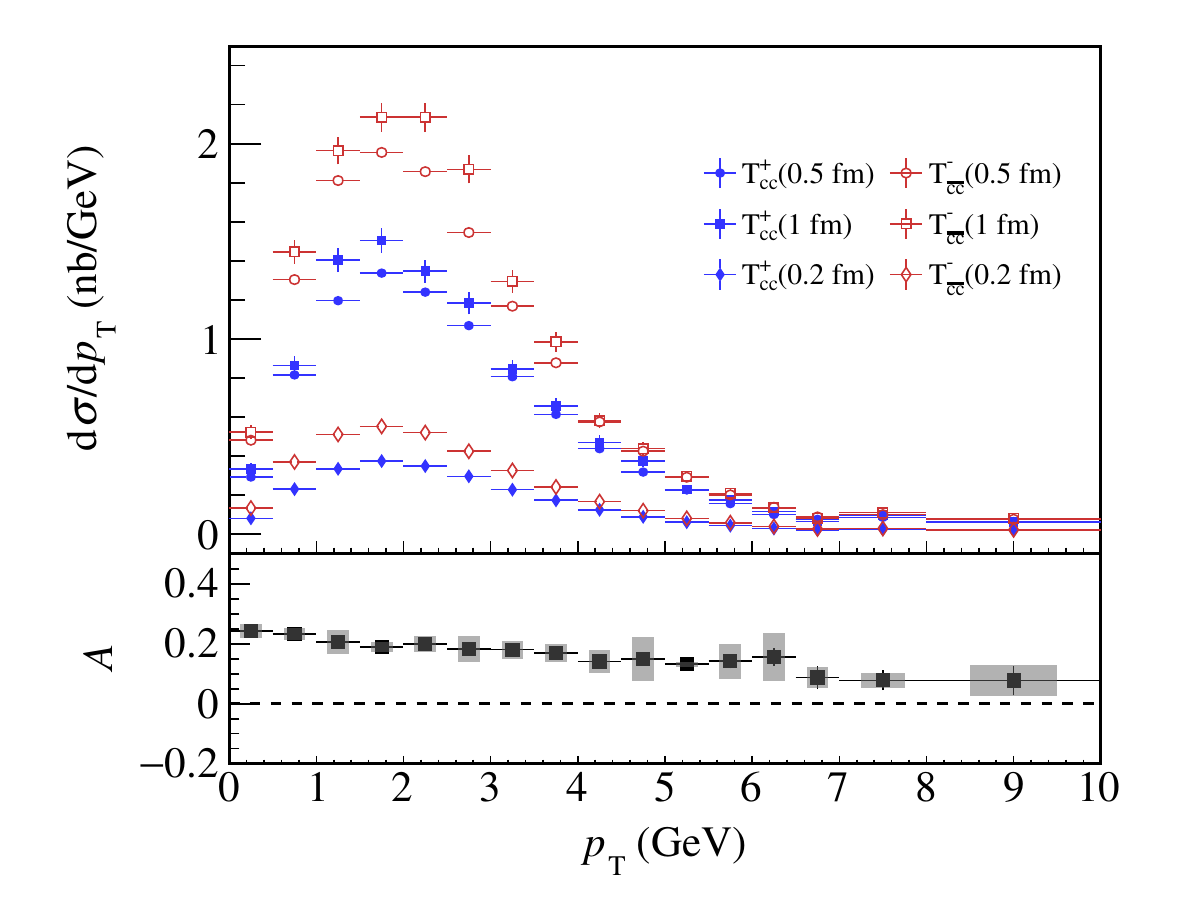}
\end{minipage}
}
\subfigure{
\begin{minipage}[b]{.9\linewidth}
\includegraphics[scale=0.4]{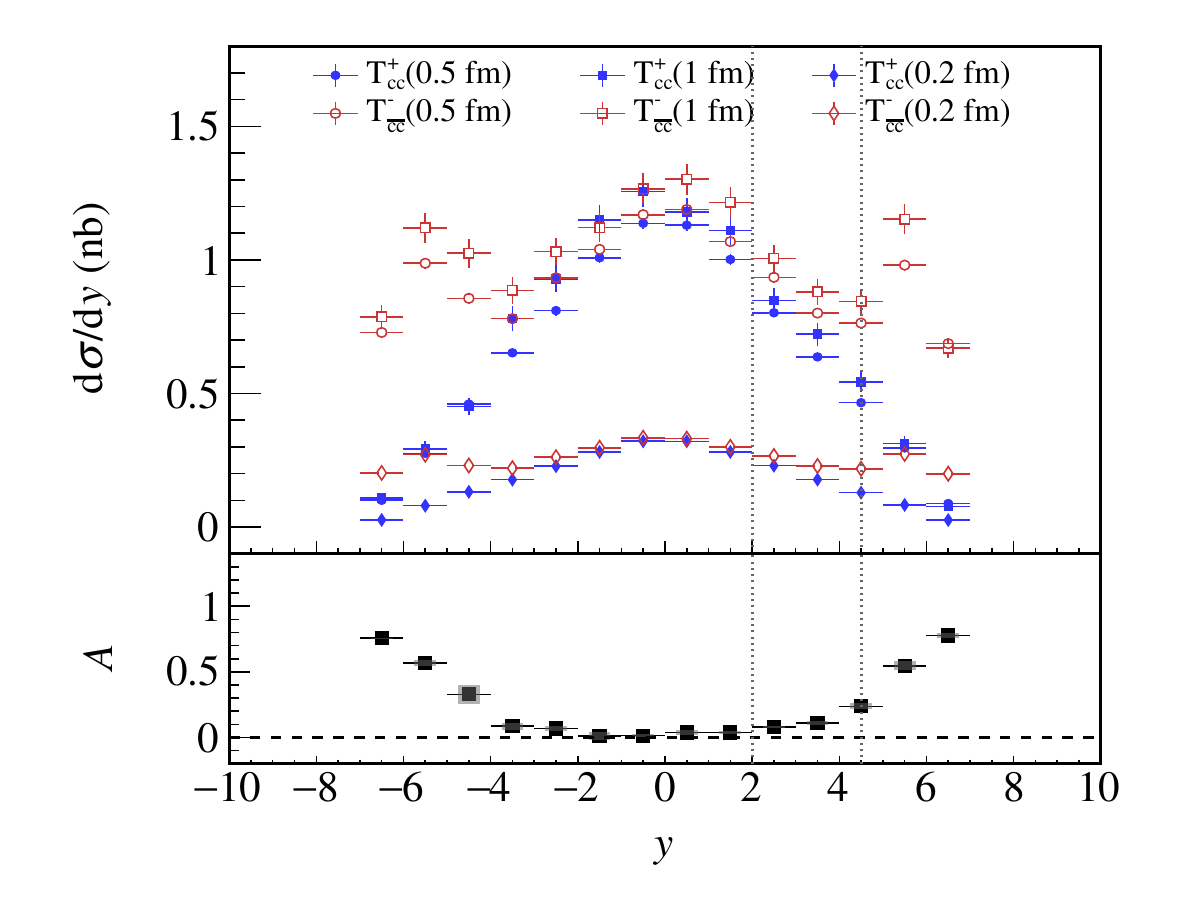}
\end{minipage}
}
\caption{\label{fig:Tetra} Similar to Fig.~\ref{fig:HM}, but for the compact tetraquark picture. The diamonds, circles, and squares correspond to the mean square radius $0.2~\mathrm{fm}$, $0.5~\mathrm{fm}$, and $1~\mathrm{fm}$, respectively.}
\end{figure}

\begin{figure}[htbp]
\centering
\includegraphics[scale=0.4]{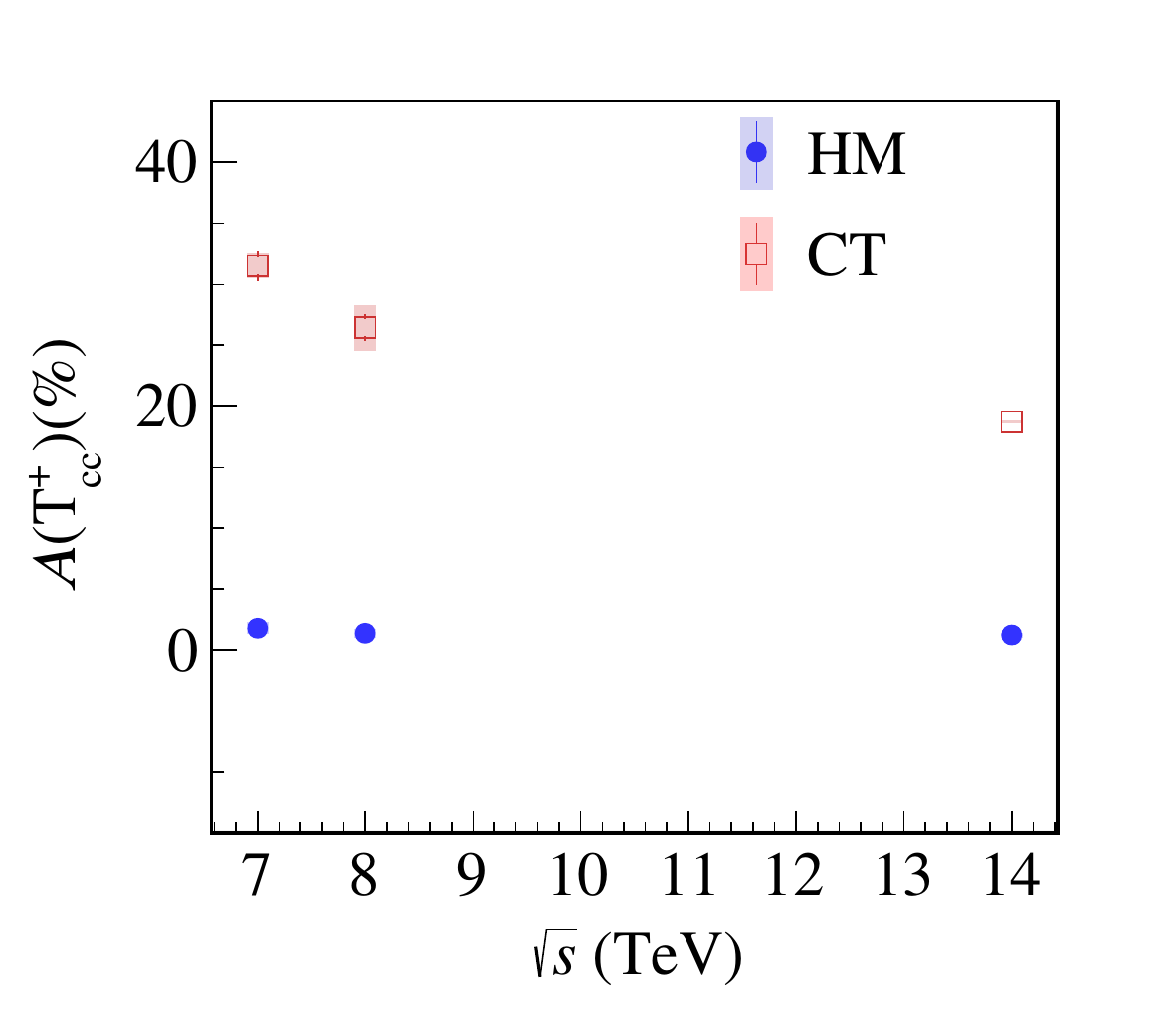}
\caption{The center-of-mass energy dependence of the asymmetry $\mathcal{A}$, defined in Eq.~\eqref{eq:A}, for the $T_{cc}^{+}$. The blue solid circles and red open squares are for the molecular (HM) and compact tetraquark (CT) pictures, respectively. The vertical bars and shaded bands represent the statistic and systematic uncertainties, respectively.}
 \label{fig:asymmetry}
\end{figure}

In the compact tetraquark picture, we simulated $100$ billion minimum-bias $pp$ collisions at $\sqrt{s} = 14~\mathrm{TeV}$.
The analogous $p_\mathrm{T}$ and $y$ distributions in the compact tetraquark picture are shown in Fig.~\ref{fig:Tetra}. 
Their cross sections are about one order of magnitude smaller than those 
of the hadronic molecular picture, as reflected in Tab.~\ref{table:cs-CT1}, based on the selection condition in the previous paragraph. 
 Although it is an order-of-magnitude estimate, the predicted transverse momentum $p_\mathrm{T}$ and rapidity $y$ distributions provide important information for further measurements. Their $p_\mathrm{T}$ distributions also reach their maximum values
at $p_\mathrm{T}\approx 2~\mathrm{GeV}$ and the $y$ distributions are 
symmetric. The systematic uncertainties are obtained by varying the size in the wave function, i.e. $0.2\sim 1~\mathrm{fm}$,
 which are strongly correlated for particle and antiparticle. One can see a significant particle and antiparticle asymmetry for the compact $T_{cc}^{+}$. Namely, the cross section of the $T_{\bar{c} \bar{c}}^{-}$ is significantly larger than that of the $T_{cc}^{+}$ at $p_\mathrm{T}\approx 2~\mathrm{GeV}$ and $y\approx 6$. The apparent difference in the low $p_\mathrm{T}$ region is due to the dominance of valence quark degrees of freedom, as indicated by the parton distribution function~\cite{ParticleDataGroup:2022pth}. The difference at large rapidity region can be attributed to the fact that the residual $ud$ diquarks within the proton predominantly travel in the beam direction and readily combine with $\bar{c} \bar{c}$ to generate $T_{\bar{c} \bar{c}}^-$. Such a feature can be quantified by the asymmetry introduced in Eq.\eqref{eq:A}.
Given the requirement for a substantially larger statistical data set to perform differential measurements of $p_T$ and $y$ distributions, the asymmetry quantity serves as a highly practical physical metric for characterizing the nature of the double charm tetraquark. In Fig.~\ref{fig:asymmetry}, it is evident that the asymmetry for the compact tetraquark picture is greater than that of the molecular picture at all three investigated energies~\footnote{ In addition to diquark and antidiquark tetraquark picture, there could also be $(c\bar{u})_8(c\bar{d})_8$ picture with $c\bar{u}$ and $c\bar{d}$ in color octet ~\cite{Ma:2023int,Yang:2009zzp,Deng:2022cld,Vijande:2009ac,Deng:2021gnb}. In this picture, the total cross sections for the $T_{cc}^+$ 
and $T_{\bar{c}\bar{c}}^-$ are $2.79\pm 0.04~\mathrm{nb}$ and $3.26 \pm 0.04~\mathrm{nb}$, respectively. The corresponding asymmetry is $\mathcal{A}=(7.91\pm 0.94)\%$ which is smaller than that of the diquark antidiquark picture, but still larger than that of the hadronic molecular picture. As the result, we conclude that 
the larger asymmetry for compact tetraquark picture is model independent.}. 
 This occurs because, 
in the hadronic molecular framework, the $T_{cc}^+$ and $T_{\bar{c}\bar{c}}^-$ are formed after hadronization, making the phase space of the constituents taken into account.

In contrast, in the compact tetraquark picture, they are formed is in the quark level. One can notice that the asymmetry $\mathcal{A}$ is a positive value for both molecular and compact tetraquark pictures, which means the total cross sections of the $T_{\bar{c} \bar{c}}^-$ are enhanced comparing to those of the $T_{cc}^+$. The uncertainties of the asymmetry $\mathcal{A}$ is obtained in the same way as that of the molecular picture.
 To check the effect of wave function for compact tetraquark, we compare the results with a Gaussian wave function (Eq.~\eqref{eq:wf1}) and a Sturmian one (Eq.~\eqref{eq:wf2}) in Tab.~\ref{table:cs-CT1} with the relative three-momentum between the two quarks of diquark less than $1.0~\mathrm{GeV}$. The detailed comparison of other cases can be found in App.~\ref{sec:app}.
 In Tab.~\ref{table:cs-CT2}, one can see that the changes of both the cross sections and the asymmetries are small. That is because the behaviors of $S$-wave functions are similar, i.e. decreasing dramatically with the increasing relative three momentum. The significant changes of CMS and ATLAS kinematic regions stem from insufficient statistics. On the contrary, the selection condition significantly modify the cross sections, making them reduced to about one half from $1~\mathrm{GeV}$ (Tab.~\ref{table:cs-CT1}) to $0.5~\mathrm{GeV}$ (Tab.~\ref{table:cs-CT3}). Although the cross sections have a factor of two changes, the asymmetry of total cross section is stable. Those for the four experimental kinematic regions are consistent within one standard deviation.
The relative systematic uncertainties of the asymmetry are significant smaller than those of the cross sections, which is because those uncertainties for particles and antiparticles are strongly correlated with each other in the asymmetry. This suggests that even in the presence of potentially large systematic uncertainties in the total cross sections resulting from parameter choices, the determination of the asymmetry remains reliable.
 Furthermore, it indicates that the asymmetry decreases with the increasing center-of-mass energy for both theoretical frameworks. This reduction is attributed to the fact that at higher energies, the predominant process is the creation of quark-antiquark pairs via energetic gluons, resulting in roughly equal production of quarks and antiquarks~\cite{ParticleDataGroup:2022pth}.
  We also present this asymmetry for various kinematic regions in the lower panels in Fig.~\ref{fig:HM} and Fig.~\ref{fig:Tetra}. One can see that the asymmetries of compact tetraquark picture for $p_\mathrm{T}<10~\mathrm{GeV}$ and each rapidity region are positive. In the hadronic molecular picture, when $p_\mathrm{T}>5~\mathrm{GeV}$, the asymmetry is consistent with zero considering the large uncertainties. Such kind of asymmetry for compact tetraquark picture would also be expected for its analog in Heavy Antiquark-Diquark Symmetry (HADS) \cite{Savage:1990di}, for instance $\Lambda_c$ or $\Lambda_b$. However their current measurement locates at the kinematic region $p_\mathrm{T}>10~\mathrm{GeV}$ \cite{CMS:2012wje} and does not have sufficient statistic to obtain a clear conclusion.

In order to harness the capabilities of the LHCb, CMS, and ATLAS detectors for assessing this asymmetry, the total cross sections for the $T_{cc}^+$, the $T_{\bar{c} \bar{c}}^-$ and their asymmetry are estimated, as shown in Tab.~\ref{table:cs-HM} and Tab.~\ref{table:cs-CT1}. It is worth noting that the cross sections in the hadronic molecular and compact tetraquark scenarios exhibit hundreds $\mathrm{nb}$ and several $\mathrm{nb}$, respectively~\footnote{ The cross section of the $T_{cc}^+$ in the hadronic molecular picture is of the same order as that in Ref.~\cite{Jin:2021cxj},
which also points out that the momentum correlation of the $DD^*$ can be viewed as a smoking gun for its internal structure.
 The cross section of the $T_{cc}^+$ in the compact tetraquark picture is significant different from that in Ref.~\cite{Qin:2020zlg},
 where only the light quarks from vacuum are considered.}. 
Additionally, the cross sections and asymmetries of the double (anti)charm tetraquarks for various kinematic regions, particularly within the LHCb ($2<y<4.5$), CMS ($|y|<1.2$), ATLAS ($|y|<0.75$) acceptance regions, are determined, as detailed in Tab.~\ref{table:cs-HM} and Tab.~\ref{table:cs-CT1}. This asymmetry for various 
kinematic regions is analogous to that defined in Ref.~\cite{Braaten:2002yt} for one single charmed meson. From the table, one can see that the LHCb experiment 
can see significantly larger asymmetry for the compact tetraquark picture than that for the molecular picture, which is sufficient to distinguish these two scenarios. As the statistics in the CMS and ATLAS kinematic regions are very small, one cannot make a clear conclusion on the asymmetry. 

That results are for $pp$ collisions at $\sqrt{s} = 14~\mathrm{TeV}$, with the largest asymmetry occurring around $|y|\sim 6$. As the center-of-mass energy decreases, the asymmetry becomes larger (Fig.~\ref{fig:asymmetry}) and the largest asymmetry occurs at a relatively smaller rapidity, which perfectly fits the LHCb acceptance region.
For the region of $p_\mathrm{T}>10~\mathrm{GeV}$, 
the cross sections of both $T_{cc}^+$ and its antiparticle are very small and exhibit substantial statistical uncertainties, which results in significant uncertainties in the asymmetry measurement.
In total, considering the current statistical data provided by the LHCb experiment,
it is indeed feasible to analyze the asymmetry of the double charm tetraquark $T_{cc}^+$, providing valuable insights into its property.

\begin{table*}[htbp]
\centering
\caption{The cross sections of the $T_{cc}^{+}$ (without brackets), $T_{\bar{c} \bar{c}}^{-}$ (with brackets) and their asymmetry $\mathcal{A}$ (the fifth column) in the compact tetraquark picture in $pp$ collisions at $\sqrt{s} = 14~\mathrm{TeV}$. The second, third and fourth column represent three different choices of the size $r$ of the compact tetraquark (Eq.~\eqref{eq:r}), i.e. $0.2~\mathrm{fm}$, $0.5~\mathrm{fm}$ and $1~\mathrm{fm}$.
The second row represents the cross sections and the asymmetry without any kinematic constraint, while the third, fourth, and fifth rows are for those with the LHCb, CMS and ATLAS kinematic acceptance, respectively. Here the diquark and antidiquark pairs are selected by requiring the relative three-momentum between the two (anti)quarks for diquark (antidiquark) less than $1~\mathrm{GeV}$, and calculated with a Gaussian wave function (Eq.~\eqref{eq:wf1}). The first and second uncertainties of the asymmetry are the systematic and statistic ones, respectively.}

\label{table:cs-CT1}
\begin{tabular}{m{11em}  m{3cm}  m{3cm} m{3cm} m{3cm}} 
  \hline\hline
  \multirow{2}{11em}{Range($\mathrm{GeV}$)} & \multicolumn{3}{c}{\multirow{2}*{$\sigma_{T_{cc}^{+}}$($\sigma_{T_{\bar{c} \bar{c}}^{-}}$)}} & \multirow{2}{9em}{$\mathcal{A}(\%)$} \\
   \\ \hline
  &\makecell*[c]{$r=0.2~\mathrm{fm}$}&\makecell*[c]{$r=0.5~\mathrm{fm}$}&\makecell*[c]{$r=1~\mathrm{fm}$}\\

  \hline\hline
  
  \multirow{3}{11em}{Full} &
  \makecell*[c]{$1.25\pm$0.005 $\mathrm{nb}$} & \makecell*[c]{$4.43\pm$0.02 $\mathrm{nb}$} & \makecell*[c]{$4.88\pm$0.02 $\mathrm{nb}$} & \multirow{3}{9em}{$18.73\pm$$0.25\pm$0.14 }\\ &
  \makecell*[c]{($1.82\pm$0.01 $\mathrm{nb}$)} & \makecell*[c]{($6.46\pm$0.02 $\mathrm{nb}$)} & \makecell*[c]{($7.16\pm$0.02 $\mathrm{nb}$)}\\ 
  
  \hline
  \specialrule{0em}{4pt}{0pt}
  \multicolumn{5}{c}{LHCb ($2<y<4.5$)}\\
  \specialrule{0em}{2pt}{0pt}

  \multirow{2}{11em}{$4<p_T<20$~\cite{LHCb:2021ten}} &
  \makecell*[c]{$39.75\pm$0.89 $\mathrm{pb}$} & \makecell*[c]{$139.88\pm$3.11 $\mathrm{pb}$} & \makecell*[c]{$163.77 \pm$3.65 $\mathrm{pb}$} & \multirow{3}{9em}{$7.35\pm$$1.48\pm$5.24 }\\ &
  \makecell*[c]{($50.57\pm$1.00 $\mathrm{pb}$)} & \makecell*[c]{($171.16\pm$3.38 $\mathrm{pb}$)} & \makecell*[c]{($163.83\pm$3.23  $\mathrm{pb}$)}\\ 
  
  \specialrule{0em}{5pt}{0pt}
  \multirow{2}{11em}{$p_T>0$~\cite{LHCb:2021auc}} &
  \makecell*[c]{$0.24 \pm$0.002 $\mathrm{nb}$} & \makecell*[c]{$0.84\pm$0.01 $\mathrm{nb}$} & \makecell*[c]{ $0.91\pm$0.01 $\mathrm{nb}$} & \multirow{3}{9em}{$11.42\pm$$0.60\pm$0.17}\\ &
  \makecell*[c]{($0.30\pm$0.002 $\mathrm{nb}$)} & \makecell*[c]{($1.05\pm$0.01 $\mathrm{nb}$)} & \makecell*[c]{($1.14\pm$ 0.01 $\mathrm{nb}$)} \\ 
  
  \hline
  \specialrule{0em}{4pt}{0pt}
   \multicolumn{5}{c}{CMS ($|y|<1.2$)} \\
   \specialrule{0em}{2pt}{0pt}
  \multirow{2}{11em}{$10<p_T<50(30)$~\cite{CMS:2013fpt}} &
  \makecell*[c]{$3.77\pm$0.51 $\mathrm{pb}$} & \makecell*[c]{$4.73\pm$0.56 $\mathrm{pb}$} & \makecell*[c]{$4.51\pm$0.53 $\mathrm{pb}$} & \multirow{3}{9em}{$-6.62\pm$$8.86\pm$7.96}\\ &
  \makecell*[c]{($1.09\pm$0.15 $\mathrm{pb}$)} & \makecell*[c]{($3.77\pm$0.51 $\mathrm{pb}$)} & \makecell*[c]{($4.94\pm$0.67 $\mathrm{pb}$)}\\ 
  \hline
  \specialrule{0em}{4pt}{0pt}
   \multicolumn{5}{c}{ATLAS ($|y|<0.75$)} \\
   \specialrule{0em}{2pt}{0pt}
   \multirow{2}{11em}{$10<p_T<70$~\cite{ATLAS:2016kwu}} &
  \makecell*[c]{$0.92\pm$0.14 $\mathrm{pb}$} & \makecell*[c]{$3.15\pm$0.46 $\mathrm{pb}$}  & \makecell*[c]{$3.11\pm$0.46 $\mathrm{pb}$ } & \multirow{3}{9em}{$0.98\pm$$11.04\pm$15.37}\\ &
  \makecell*[c]{($0.69\pm$0.12 $\mathrm{pb}$)} & \makecell*[c]{($2.83\pm$0.49 $\mathrm{pb}$)} & \makecell*[c]{($4.88\pm$0.84 $\mathrm{pb}$)} \\ 
  
  \hline\hline
\end{tabular}
\end{table*}

\vspace{0.2cm}
\section{Summary}

We propose a direct measurable physical quantity, the production asymmetry of particles and antiparticles, to probe the nature of doubly heavy exotic candidates in $pp$ collision. The asymmetry is defined as $\mathcal{A}\equiv ({\sigma^--\sigma^+})/({\sigma^-+\sigma^+})$ where $\sigma^+$ and $\sigma^-$ represent the cross sections of particle and antiparticle, respectively. Taking the double charm tetraquark $T_{cc}^+$ as an illustration, the asymmetry of the compact tetrquark picture is significantly larger than that of the molecular picture, making it an unambiguous quantity for distinguishing these two scenarios. In both scenarios, the asymmetries decrease with the increasing center-of-mass energy, owing to the dominant role played by gluons at high energies, which produce an equal number of antiparticles and particles. 
Furthermore, with sufficient data samples from $pp$ collisions, one can observe that the differential cross section of $T_{\bar{c} \bar{c}}^-$ is significantly larger than that of $T_{cc}^+$ at $p_\mathrm{T}\approx 2~\mathrm{GeV}$ 
and $y \approx \pm 6$ at a center-of-mass energy of 14$~\mathrm{TeV}$ in the compact tetraquark picture. The excess in the large rapidity region is a consequence of the residual $ud$ diquarks in the proton predominantly traveling in the beam direction. 
Our work can be extended to the exploration of other double heavy tetraquark candidates, offering a versatile aproach to advance our understanding of exotic hadrons.

\begin{acknowledgments}
We are grateful to Yi-Fu Cai, Meng-Lin Du, Feng-Kun Guo, Shi-Yuan Li, 
Eulogio Oset, Hao Qiu, Qin Qin, 
Cheng-Ping Shen, Li-Ming Zhang, Shan-Liang Zhang for the helpful discussion. 
This work is partly supported by the National Natural Science Foundation of China with Grant Nos.~12375073, 12275091, 12235018, and 12035007, Guangdong Provincial funding with Grant Nos.~2019QN01X172 and 2023A1515010416, Guangdong Major Project of Basic and Applied Basic Research No.~2020B0301030008,
 the NSFC and the Deutsche Forschungsgemeinschaft (DFG, German
Research Foundation) through the funds provided to the Sino-German Collaborative
Research Center TRR110 ``Symmetries and the Emergence of Structure in QCD"
(NSFC Grant No. 12070131001, DFG Project-ID 196253076-TRR 110), National Key Basic Research Program of China under Contract No. 2020YFA0406300, and Strategic Priority Research Program of Chinese Academy of Sciences (Grant No. XDB34030302).
\end{acknowledgments}

\vspace{2cm}

\appendix
\section{The cross sections and asymmetry in compact tetraquark picture } 
\label{sec:app}

In this section, we present the stability of the asymmetry, especially for the total cross sections and those within the LHCb kinematic acceptance. 

The cross sections and asymmetries in Tab.~\ref{table:cs-CT3} is similar to those in Tab.~\ref{table:cs-CT1}, but for the relative three-momentum between the two (anti)quarks of diquark(antidiquark) less than $0.5~\mathrm{GeV}$. One can see that the asymmetries for the total cross sections and LHCb kinematic regions are consistent with those in Tab.~\ref{table:cs-CT1} within one standard deviation. This indicates the selection condition of (anti)diquarks will not change the asymmetry larger than one standard deviation. 

\begin{table*}[htbp]
\centering
\caption{Similar to Tab.~\ref{table:cs-CT1}, but for the relative three-momentum between the two (anti)quarks of diquark (antidiquark) less than $0.5~\mathrm{GeV}$.}
\label{table:cs-CT3}
\begin{tabular}{m{11em}  m{3cm}  m{3cm} m{3cm} m{3cm}} 
  \hline\hline
  \multirow{2}{11em}{Range($\mathrm{GeV}$)} & \multicolumn{3}{c}{\multirow{2}*{$\sigma_{T_{cc}^{+}}$($\sigma_{T_{\bar{c} \bar{c}}^{-}}$)}} & \multirow{2}{9em}{$\mathcal{A}(\%)$} \\
   \\ \hline
  &\makecell*[c]{$r=0.2\mathrm{fm}$}&\makecell*[c]{$r=0.5\mathrm{fm}$}&\makecell*[c]{$r=1\mathrm{fm}$}\\

  \hline\hline
  
  \multirow{3}{11em}{Full} &
  \makecell*[c]{$0.23\pm$0.002 $\mathrm{nb}$} & \makecell*[c]{$2.20\pm$0.02 $\mathrm{nb}$} & \makecell*[c]{$4.72\pm$0.05 $\mathrm{nb}$} & \multirow{3}{9em}{$18.71\pm$$0.67\pm$0.17 }\\ &
  \makecell*[c]{($0.34\pm$0.003 $\mathrm{nb}$)} & \makecell*[c]{($3.21\pm$0.03 $\mathrm{nb}$)} & \makecell*[c]{($6.92\pm$0.06 $\mathrm{nb}$)}\\ 
  
  \hline
  \specialrule{0em}{4pt}{0pt}
  \multicolumn{5}{c}{LHCb ($2<y<4.5$)}\\
  \specialrule{0em}{2pt}{0pt}

  \multirow{2}{11em}{$4<p_T<20$~\cite{LHCb:2021ten}} &
  \makecell*[c]{$7.18\pm$0.44 $\mathrm{pb}$} & \makecell*[c]{$69.86\pm$4.24 $\mathrm{pb}$} & \makecell*[c]{$158.77 \pm$9.63 $\mathrm{pb}$} & \multirow{3}{9em}{$3.53\pm$$4.15\pm$2.99 }\\ &
  \makecell*[c]{($8.15\pm$0.46 $\mathrm{pb}$)} & \makecell*[c]{($77.10\pm$4.38 $\mathrm{pb}$)} & \makecell*[c]{($156.83\pm$8.91  $\mathrm{pb}$)}\\ 
  
  \specialrule{0em}{5pt}{0pt}
  \multirow{2}{11em}{$p_T>0$~\cite{LHCb:2021auc}} &
  \makecell*[c]{$0.04 \pm$0.001 $\mathrm{nb}$} & \makecell*[c]{$0.41\pm$0.01 $\mathrm{nb}$} & \makecell*[c]{ $0.88\pm$0.02 $\mathrm{nb}$} & \multirow{3}{9em}{$10.65\pm$$1.64\pm$0.42}\\ &
  \makecell*[c]{($0.05\pm$0.001 $\mathrm{nb}$)} & \makecell*[c]{($0.51\pm$0.01 $\mathrm{nb}$)} & \makecell*[c]{($1.10\pm$ 0.02 $\mathrm{nb}$)} \\ 
  
  \hline
  \specialrule{0em}{4pt}{0pt}
   \multicolumn{5}{c}{CMS ($|y|<1.2$)} \\
   \specialrule{0em}{2pt}{0pt}
  \multirow{2}{11em}{$10<p_T<50(30)$~\cite{CMS:2013fpt}} &
  \makecell*[c]{$0.24\pm$0.08 $\mathrm{pb}$} & \makecell*[c]{$2.25\pm$0.75 $\mathrm{pb}$} & \makecell*[c]{$4.34\pm$1.45 $\mathrm{pb}$} & \multirow{3}{9em}{$-9.84\pm$$25.79\pm$10.79}\\ &
  \makecell*[c]{($0.16\pm$0.07 $\mathrm{pb}$)} & \makecell*[c]{($1.65\pm$0.67 $\mathrm{pb}$)} & \makecell*[c]{($4.82\pm$1.97 $\mathrm{pb}$)}\\ 
  \hline
  \specialrule{0em}{4pt}{0pt}
   \multicolumn{5}{c}{ATLAS ($|y|<0.75$)} \\
   \specialrule{0em}{2pt}{0pt}
   \multirow{2}{11em}{$10<p_T<70$~\cite{ATLAS:2016kwu}} &
  \makecell*[c]{$0.16\pm$0.06 $\mathrm{pb}$} & \makecell*[c]{$1.51\pm$0.62 $\mathrm{pb}$}  & \makecell*[c]{$3.00\pm$1.22 $\mathrm{pb}$ } & \multirow{3}{9em}{$9.44\pm$$28.33\pm$9.92}\\ &
  \makecell*[c]{($0.16\pm$0.07 $\mathrm{pb}$)} & \makecell*[c]{($1.65\pm$0.67 $\mathrm{pb}$)} & \makecell*[c]{($4.82\pm$1.97 $\mathrm{pb}$)} \\ 
  
  \hline\hline
\end{tabular}
\end{table*}

Table~\ref{table:cs-CT2}, in comparison of Tab.~\ref{table:cs-CT1} presents the effect 
of the wave function on the asymmetry. 
The comparison between Tab.~\ref{table:cs-CT2} and Tab.~\ref{table:cs-CT4} also presents the selection condition on the asymmetry, but with Sturmian wave function. 
For all the above cases, the asymmetries are consistent with each other within one standard deviation.

\begin{table*}[htbp]
\centering
\caption{Similar to Tab.~\ref{table:cs-CT1}, but with a Sturmian wave function (Eq.~\eqref{eq:wf2}) instead of a Gaussian one.}
\label{table:cs-CT2}
\begin{tabular}{m{11em}  m{3cm}  m{3cm} m{3cm} m{3cm}} 
  \hline\hline
  \multirow{2}{11em}{Range($\mathrm{GeV}$)} & \multicolumn{3}{c}{\multirow{2}*{$\sigma_{T_{cc}^{+}}$($\sigma_{T_{\bar{c} \bar{c}}^{-}}$)}} & \multirow{2}{9em}{$\mathcal{A}(\%)$} \\
   \\ \hline
  &\makecell*[c]{$r=0.2\mathrm{fm}$}&\makecell*[c]{$r=0.5\mathrm{fm}$}&\makecell*[c]{$r=1\mathrm{fm}$}\\

  \hline\hline
  
  \multirow{3}{11em}{Full} &
  \makecell*[c]{$1.50\pm$0.01 $\mathrm{nb}$} & \makecell*[c]{$4.16\pm$0.02 $\mathrm{nb}$} & \makecell*[c]{$4.84\pm$0.02 $\mathrm{nb}$} & \multirow{3}{9em}{$18.70\pm$$0.25\pm$0.08 }\\ &
  \makecell*[c]{($2.18\pm$0.01 $\mathrm{nb}$)} & \makecell*[c]{($6.07\pm$0.02 $\mathrm{nb}$)} & \makecell*[c]{($7.09\pm$0.02 $\mathrm{nb}$)}\\ 
  
  \hline
  \specialrule{0em}{4pt}{0pt}
  \multicolumn{5}{c}{LHCb ($2<y<4.5$)}\\
  \specialrule{0em}{2pt}{0pt}

  \multirow{2}{11em}{$4<p_T<20$~\cite{LHCb:2021ten}} &
  \makecell*[c]{$47.53\pm$1.06 $\mathrm{pb}$} & \makecell*[c]{$132.57\pm$2.95 $\mathrm{pb}$} & \makecell*[c]{$159.48 \pm$3.55 $\mathrm{pb}$} & \multirow{3}{9em}{$6.71\pm$$1.48\pm$5.17 }\\ &
  \makecell*[c]{($60.31\pm$1.19 $\mathrm{pb}$)} & \makecell*[c]{($157.59\pm$3.11 $\mathrm{pb}$)} & \makecell*[c]{($158.35\pm$3.13  $\mathrm{pb}$)}\\ 
  
  \specialrule{0em}{5pt}{0pt}
  \multirow{2}{11em}{$p_T>0$~\cite{LHCb:2021auc}} &
  \makecell*[c]{$0.29 \pm$0.003 $\mathrm{nb}$} & \makecell*[c]{$0.79\pm$0.01 $\mathrm{nb}$} & \makecell*[c]{ $0.90\pm$0.01 $\mathrm{nb}$} & \multirow{3}{9em}{$11.38\pm$$0.60\pm$0.16}\\ &
  \makecell*[c]{($0.36\pm$0.003 $\mathrm{nb}$)} & \makecell*[c]{($0.99\pm$0.01 $\mathrm{nb}$)} & \makecell*[c]{($1.13\pm$ 0.01 $\mathrm{nb}$)} \\ 
  
  \hline
  \specialrule{0em}{4pt}{0pt}
   \multicolumn{5}{c}{CMS ($|y|<1.2$)} \\
   \specialrule{0em}{2pt}{0pt}
  \multirow{2}{11em}{$10<p_T<50(30)$~\cite{CMS:2013fpt}} &
  \makecell*[c]{$1.69\pm$0.20 $\mathrm{pb}$} & \makecell*[c]{$4.37\pm$0.52 $\mathrm{pb}$} & \makecell*[c]{$4.02\pm$0.47 $\mathrm{pb}$} & \multirow{3}{9em}{$-2.69\pm$$8.81\pm$12.52}\\ &
  \makecell*[c]{($1.31\pm$0.18 $\mathrm{pb}$)} & \makecell*[c]{($3.57\pm$0.48 $\mathrm{pb}$)} & \makecell*[c]{($5.44\pm$0.73 $\mathrm{pb}$)}\\ 
  \hline
  \specialrule{0em}{4pt}{0pt}
   \multicolumn{5}{c}{ATLAS ($|y|<0.75$)} \\
   \specialrule{0em}{2pt}{0pt}
   \multirow{2}{11em}{$10<p_T<70$~\cite{ATLAS:2016kwu}} &
  \makecell*[c]{$1.10\pm$0.16 $\mathrm{pb}$} & \makecell*[c]{$2.92\pm$0.43 $\mathrm{pb}$}  & \makecell*[c]{$2.79\pm$0.41 $\mathrm{pb}$ } & \multirow{3}{9em}{$5.44\pm$$10.89\pm$18.41}\\ &
  \makecell*[c]{($0.85\pm$0.15 $\mathrm{pb}$)} & \makecell*[c]{($2.83\pm$0.49 $\mathrm{pb}$)} & \makecell*[c]{($5.26\pm$0.90 $\mathrm{pb}$)} \\ 
  
  \hline\hline
\end{tabular}
\end{table*}

\begin{table*}[htbp]
\centering
\caption{Similar to Tab.~\ref{table:cs-CT2}, but for the relative three-momentum between diquark and antidiquark less than $0.5~\mathrm{GeV}$.}
\label{table:cs-CT4}
\begin{tabular}{m{11em}  m{3cm}  m{3cm} m{3cm} m{3cm}} 
  \hline\hline
  \multirow{2}{11em}{Range($\mathrm{GeV}$)} & \multicolumn{3}{c}{\multirow{2}*{$\sigma_{T_{cc}^{+}}$($\sigma_{T_{\bar{c} \bar{c}}^{-}}$)}} & \multirow{2}{9em}{$\mathcal{A}(\%)$} \\
   \\ \hline
  &\makecell*[c]{$r=0.2\mathrm{fm}$}&\makecell*[c]{$r=0.5\mathrm{fm}$}&\makecell*[c]{$r=1\mathrm{fm}$}\\

  \hline\hline
  
  \multirow{3}{11em}{Full} &
  \makecell*[c]{$0.34\pm$0.004 $\mathrm{nb}$} & \makecell*[c]{$2.39\pm$0.03 $\mathrm{nb}$} & \makecell*[c]{$4.42\pm$0.05 $\mathrm{nb}$} & \multirow{3}{9em}{$18.70\pm$$0.67\pm$0.11 }\\ &
  \makecell*[c]{($0.49\pm$0.004 $\mathrm{nb}$)} & \makecell*[c]{($3.49\pm$0.03 $\mathrm{nb}$)} & \makecell*[c]{($6.47\pm$0.06 $\mathrm{nb}$)}\\ 
  
  \hline
  \specialrule{0em}{4pt}{0pt}
  \multicolumn{5}{c}{LHCb ($2<y<4.5$)}\\
  \specialrule{0em}{2pt}{0pt}

  \multirow{2}{11em}{$4<p_T<20$~\cite{LHCb:2021ten}} &
  \makecell*[c]{$10.48\pm$0.64 $\mathrm{pb}$} & \makecell*[c]{$76.95\pm$4.67$\mathrm{pb}$} & \makecell*[c]{$146.22 \pm$8.87 $\mathrm{pb}$} & \multirow{3}{9em}{$2.59\pm$$4.15\pm$3.43 }\\ &
  \makecell*[c]{($11.83\pm$0.67 $\mathrm{pb}$)} & \makecell*[c]{($83.02\pm$4.72 $\mathrm{pb}$)} & \makecell*[c]{($140.27\pm$7.97  $\mathrm{pb}$)}\\ 
  
  \specialrule{0em}{5pt}{0pt}
  \multirow{2}{11em}{$p_T>0$~\cite{LHCb:2021auc}} &
  \makecell*[c]{$0.06 \pm$0.002 $\mathrm{nb}$} & \makecell*[c]{$0.45\pm$0.01 $\mathrm{nb}$} & \makecell*[c]{ $0.82\pm$0.02 $\mathrm{nb}$} & \multirow{3}{9em}{$10.69\pm$$1.64\pm$0.36}\\ &
  \makecell*[c]{($0.08\pm$0.002 $\mathrm{nb}$)} & \makecell*[c]{($0.56\pm$0.01 $\mathrm{nb}$)} & \makecell*[c]{($1.03\pm$ 0.02 $\mathrm{nb}$)} \\ 
  
  \hline
  \specialrule{0em}{4pt}{0pt}
   \multicolumn{5}{c}{CMS ($|y|<1.2$)} \\
   \specialrule{0em}{2pt}{0pt}
  \multirow{2}{11em}{$10<p_T<50(30)$~\cite{CMS:2013fpt}} &
  \makecell*[c]{$0.34\pm$0.11 $\mathrm{pb}$} & \makecell*[c]{$2.41\pm$0.80 $\mathrm{pb}$} & \makecell*[c]{$3.57\pm$1.19 $\mathrm{pb}$} & \multirow{3}{9em}{$-4.32\pm$$25.67\pm$15.52}\\ &
  \makecell*[c]{($0.24\pm$0.10 $\mathrm{pb}$)} & \makecell*[c]{($1.91\pm$0.78 $\mathrm{pb}$)} & \makecell*[c]{($5.06\pm$2.06 $\mathrm{pb}$)}\\ 
  \hline
  \specialrule{0em}{4pt}{0pt}
   \multicolumn{5}{c}{ATLAS ($|y|<0.75$)} \\
   \specialrule{0em}{2pt}{0pt}
   \multirow{2}{11em}{$10<p_T<70$~\cite{ATLAS:2016kwu}} &
  \makecell*[c]{$0.23\pm$0.09 $\mathrm{pb}$} & \makecell*[c]{$1.63\pm$0.66 $\mathrm{pb}$}  & \makecell*[c]{$2.49\pm$1.02 $\mathrm{pb}$ } & \multirow{3}{9em}{$14.44\pm$$27.69\pm$14.09}\\ &
  \makecell*[c]{($0.24\pm$0.10 $\mathrm{pb}$)} & \makecell*[c]{($1.91\pm$0.78 $\mathrm{pb}$)} & \makecell*[c]{($5.06\pm$2.06 $\mathrm{pb}$)} \\ 
  
  \hline\hline
\end{tabular}
\end{table*}

\clearpage

\end{document}